\newif\ifmainmode        \mainmodetrue
\newif\ifsuppmode         \suppmodefalse
\newif\ifcombinedmode \combinedmodefalse

\documentclass[
aps,
prx,
showpacs,
preprintnumbers,
twocolumn,
nosuperscriptaddress,
nofootinbib,
10pt,
english,
longbibliography
]{revtex4-2}

\usepackage{physics}
\usepackage{amsmath}
\usepackage{mathtools}
\usepackage{dsfont}

\usepackage[T1]{fontenc}
\usepackage{newtxtext}
\usepackage{newtxmath}

\usepackage{enumitem}

\usepackage{appendix}

\usepackage{placeins}

\usepackage{color}
\usepackage[dvipsnames]{xcolor}

\definecolor{nqdcolor}{rgb}{0.5586, 0.0586, 0.4219}

\usepackage{hyperref}
\hypersetup{colorlinks,citecolor=nqdcolor,linkcolor=nqdcolor,urlcolor=nqdcolor}

\usepackage{makecell}
\usepackage{colortbl} 

\usepackage{graphicx}
\usepackage[all]{hypcap}
\graphicspath{{../visual_elements/figs/}{../visual_elements/tables/}{../visual_elements/videos/}}

\newcommand{\phdagger}[0]{{\phantom{\dagger}}}

\newcommand{\nocontentsline}[3]{}
\let\origcontentsline\addcontentsline
\newcommand\stoptoc{\let\addcontentsline\nocontentsline}
\newcommand\resumetoc{\let\addcontentsline\origcontentsline}

\makeatletter
\@ifundefined{ifmainmode}{\newif\ifmainmode\mainmodetrue}{}%
\@ifundefined{ifsuppmode}{\newif\ifsuppmode\suppmodetrue}{}%
\@ifundefined{ifcombinedmode}{\newif\ifcombinedmode\combinedmodetrue}{}%
\makeatother

\begin{document}



\author{Federico Balducci}
\email{fbalducci@pks.mpg.de}
\affiliation{Max Planck Institute for the Physics of Complex Systems, N\"othnitzer Str.\ 38, 01187 Dresden, Germany}

\author{Paul M. Schindler}
\email{psch@pks.mpg.de}
\affiliation{Max Planck Institute for the Physics of Complex Systems, N\"othnitzer Str.\ 38, 01187 Dresden, Germany}

\author{Andrea Solfanelli}
\email{solfanelli@pks.mpg.de}
\affiliation{Max Planck Institute for the Physics of Complex Systems, N\"othnitzer Str.\ 38, 01187 Dresden, Germany}

\author{Marin Bukov}
\affiliation{Max Planck Institute for the Physics of Complex Systems, N\"othnitzer Str.\ 38, 01187 Dresden, Germany}

\title{(Non-)Traversable Quantum Phase Transitions}

\ifmainmode

\begin{abstract}
    Quantum phase transitions manifest as an abrupt change in the ground state of a many-body system; yet it is an open question whether this sudden change necessarily precludes a continuous dynamical connection between the two phases.
    We introduce a classification of quantum phase transitions based on this geometric aspect of the ground-state manifold, that differs from known classifications.
    By leveraging the framework of counterdiabatic driving, we explicitly construct schedules that dynamically connect one phase to another.
    This strategy allows us to uncover a large class of quantum phase transitions, where the states on both sides are separated only by a finite geometric distance in the thermodynamic limit.
    We term such transitions \emph{traversable}, since exact counterdiabatic driving links the two phases via a finite dynamical protocol in the thermodynamic limit.
    We show that multiple known transitions fall into this class---e.g., symmetry-breaking transitions obeying hyperscaling and discontinuous transitions with an enhanced continuous symmetry.
    We further show the existence of quantum phase transitions that cannot be crossed dynamically even with the help of nonlocal counterdiabatic driving, as they would require divergent amplitudes and frequencies.
    Geometrically, these \emph{nontraversable} transitions correspond to an infinite distance separating the two phases of matter; we show that the class comprises continuous transitions exhibiting mean-field universality, and discontinuous transitions arising from the competition between metastable minima.
    Our geometric classification goes beyond the known taxonomy, is independent of local order parameters and renormalization group fixed points, and has direct implications for the complexity of state preparation and adiabatic quantum computation.
\end{abstract}

\maketitle
	

\section{Introduction} 
\label{sec:intro}

\begin{figure*}[t]
    \centering
    \includegraphics[width=\linewidth]{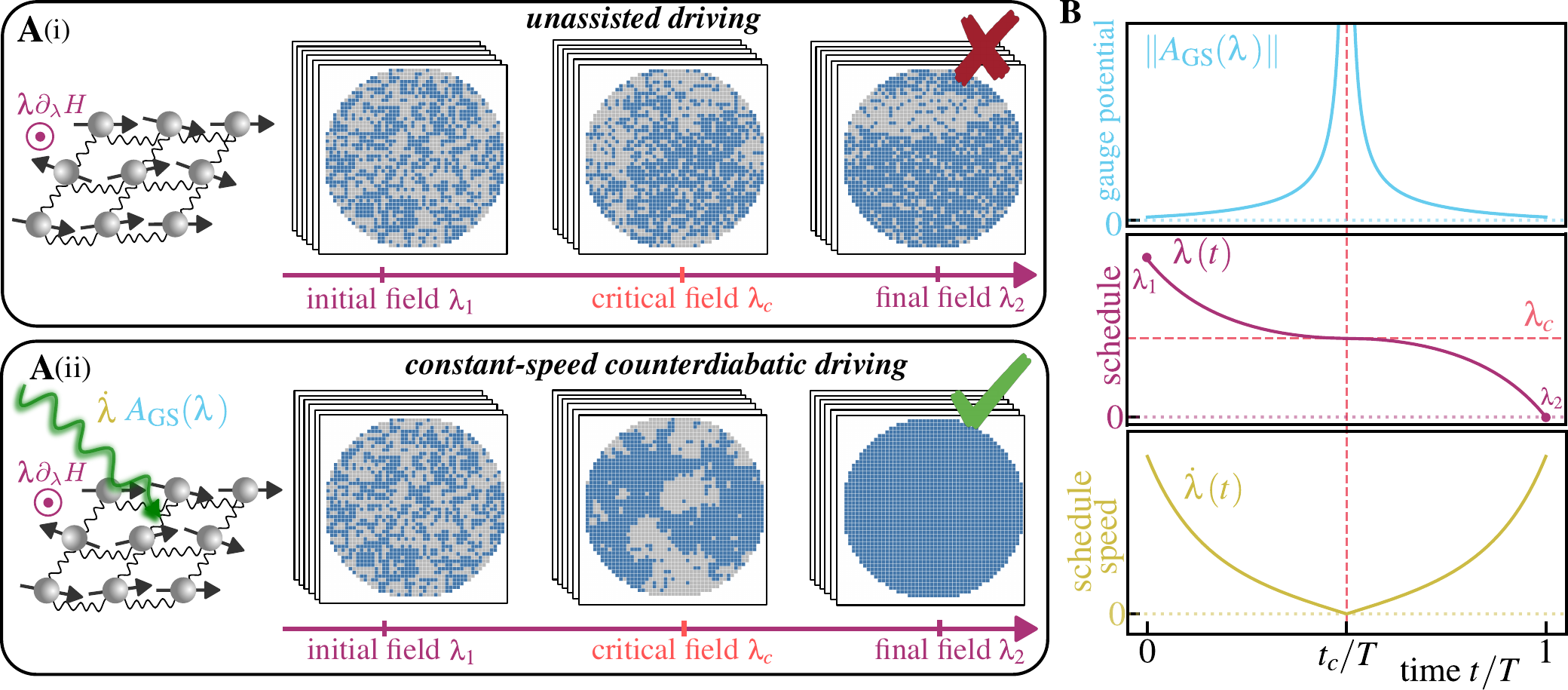}
    \caption{
        \textbf{Constant-speed protocol and traversability of quantum phase transitions} 
        \textbf{A}, in this work we consider groundstate manipulation by driving along control parameter $\lambda$ from $\lambda_1$ to $\lambda_2$ through a phase transitions at $\lambda=\lambda_c$: (i) unassisted driving $H(\lambda)$ leads to excitations in the final state; using the constant-speed counterdiabatic driving schedule, $H + \dot{\lambda}_\text{cs} A_\text{GS}$, transitionless driving can be achieved.
        \textbf{B}, amplitude of gauge potential $A_\text{GS}(\lambda)$~(blue) diverges at the critical point. The idea of the constant-speed protocol is to use a singular driving schedule $\lambda_\text{CS}$ (red)---with speed $\dot{\lambda}=0$ (orange) at the critical point---which cancels the gauge potential's divergence; consequently, the quantum phase transition can be traversed with a finite norm counterdiabatic driving Hamiltonian $\norm{\dot{\lambda} A_\text{GS}}{<}\infty$.
    }
    \label{fig:CD-mechanism}
\end{figure*}

When many degrees of freedom interact, qualitatively new phenomena arise that cannot be inferred from individual behavior alone---a perspective famously expressed by P.~W.~Anderson~\cite{anderson1972more}. 
Their macroscopic description requires emergent variables and principles different from the microscopic laws from which they arise. 
Phase transitions and critical phenomena stand as a paradigmatic example---a cornerstone of statistical physics in the study of emergent collective behavior~\cite{fisher1998renormalization}. 
Familiar instances range from the freezing or boiling processes in liquids to the formation of magnets and crystals, but can extend to far less intuitive phenomena, such as the condensation of electrons in superconductors or of atoms into superfluids~\cite{goldenfeld2018lectures,Sachdev2011Quantum}. 

In quantum systems at zero temperature, this intrinsic emergence of macroscopic complexity is constrained by an underlying geometric structure~\cite{Kolodrubetz2017Geometry,Carollo2020Geometry,strelecek2025quantum}. 
At the (quantum) phase transition, the collective macroscopic behavior changes abruptly as a result of the system crossing a geometric boundary that forbids the smooth continuation of its state.
We show that these singular changes in the geometry of state space can serve as an organizing principle for macroscopic phases of matter.

Over the years, different classifications of phase transitions have been developed to systematize the observed abundant complexity:
a continuous (second-order)/discontinuous (first-order) taxonomy considers the behavior of the order parameter at the boundary of the ordered phase;
an equilibrium/nonequilibrium distinction refers to the breaking of the microscopic principle of detailed balance;
a thermal/quantum axis is based on the origin of the underlying fluctuations triggering the transition~\cite{goldenfeld2018lectures}.
Here, we propose a dichotomous classification of quantum phase transitions (QPT) based on the geometry of the ground state manifold.

Whereas most known classifications consider equilibrium features, the latter can be probed by analyzing the excitations created by nonequilibrium drives. A classic example is the Kibble-Zurek mechanism for crossing critical points~\cite{Kibble1976Topology,*Kibble1980Some,Zurek1985Cosmological,Zurek2005Dynamics,delCampo2014Universality}: due to the closing of the ground-state energy gap at the transition point, adiabatic evolution breaks down, exciting the system. The density of these excitations exhibits a characteristic scaling with the distance from the critical point, and retains measurable information about the critical properties of the transition~\cite{Zurek1985Cosmological,Bauerle1996Laboratory,Ducci1999Order,Carmi2000Observation,Chae2012Direct,Lamporesi2013Spontaneous,delCampo2014Universality,Deutschlander2015Kibble}.
Thus, the Kibble-Zurek mechanism reveals equilibrium properties of critical points via departure from adiabatic evolution.

By contrast, rather than utilizing excitations to learn about phase transition properties, here we suppress them completely using transitionless counterdiabatic (CD) driving, which generates perfect adiabatic evolution away from the adiabatic regime~\cite{Demirplak2003Adiabatic,*Demirplak2005Assisted,*Demirplak2008Consistency,Berry2009Transitionless,Jarzynski2013Generating}, see Fig.~\ref{fig:CD-mechanism}A. This is achieved by implementing additional forces, described by the adiabatic gauge potential (AGP)~\cite{Kolodrubetz2017Geometry}, that cancel exactly the occurrence of any excitations.
CD evolution is inherently related to parallel transport on the ground-state manifold, whose geometric invariants exhibit singularities at the QPT point~\cite{Kolodrubetz2013Classifying}.

Counterdiabatic driving across critical points has been studied analytically for a handful of exactly solvable models~\cite{delCampo2012Assisted,Takahashi2013Transitionless,Grabarits2026Fighting}, and numerically for generic interacting systems away from the thermodynamic regime~\cite{Sels2017Minimizing,Grabarits2026Fighting}. 
A major conceptual challenge is the divergence of the adiabatic gauge potential at the QPT point, see Fig.~\ref{fig:CD-mechanism}B. 
This leads to the breakdown of CD driving, which is often misinterpreted as a consequence of the failure of adiabatic evolution across the phase transition, and conflated with the loss of locality of the adiabatic gauge potential. 

Placing a clear distinction between these two notions, we demonstrate that a large class of quantum phase transition points can be traversed with unit fidelity following transitionless counterdiabatic evolution in the thermodynamic limit. We achieve this by constructing a drive schedule that slows down at the phase transition point faster than the nonlocal gauge potential diverges~\cite{Holtzman2025Shortcuts}, cf.~Fig.~\ref{fig:CD-mechanism}B.
Crucially, the control schedule has a well-defined functional form in the thermodynamic limit, i.e.,\ it does not feature divergent amplitude or frequency scales as the system size increases. 
This allows us to identify a thermodynamic regime that enables us to probe the underlying physics in numerical simulations with finite resources.   
Whereas the schedule requires access to the exact ground-state gauge potential, which grows nonlocal at the QPT, the instantaneous generator of the unitary, connecting the ground states on both sides of the phase transition, has a finite norm density; as a result, the evolution can be implemented using a finite-depth circuit with multi-body operations.

Rather more surprisingly, we also find that the inherent structure of many QPTs precludes the existence of finite unit-fidelity CD schedules, even when access to the exact ground-state gauge potential is granted. Thus, we propose a universal dichotomous classification of quantum phase transitions as \textit{traversable} or \textit{nontraversable}, summarized in Fig.~\ref{fig:QPT-tree}.
Examples of traversable QPTs include all phase transitions that obey hyperscaling, as defined by the theory of critical phenomena~\cite{fisher1998renormalization}: for instance, certain symmetry-breaking phase transitions (e.g.,\ ferromagnet-to-paramagnet in local quantum Ising chains), and some symmetry-protected topological phase transitions, as we explicitly discuss. Discontinuous transitions with enhanced continuous symmetry can also be traversable. 
We demonstrate that traversability is a dynamical, geometric property that is not related to the existence of a local order parameter, and is not confined to fixed points of the renormalization group flow (criticality); it goes beyond the existing taxonomy of known classifications.
Examples of nontraversable QPTs are ubiquitously encountered in models in the mean-field universality class (e.g.,\ the ferromagnetic-to-paramagnetic transition in the Ising Lipkin-Meshkov-Glick model, or the superradiance transition in the Dicke model), as well as systems exhibiting discontinuous phase transitions with exponentially closing energy gaps; the latter have no-go implications for CD-assisted adiabatic quantum computing.  

We show that, at its core, traversability can be inferred from the geometric properties of the ground state manifold in projective Hilbert space. In particular, traversable QPTs are those separating two phases connected by a finite-length path. As a consequence, the ground state manifold admits a continuous parametrization and the ground state changes continuously across the critical point. By contrast, for nontraversable QPTs, the ground state manifold has an intrinsic singularity at the QPT point that cannot be removed by reparametrizing the coordinate chart.  
Our work emphasizes the central role played by geometry as an organizing principle for interacting quantum matter.

\begin{figure*}[t]
    \centering
    \includegraphics[width=\linewidth]{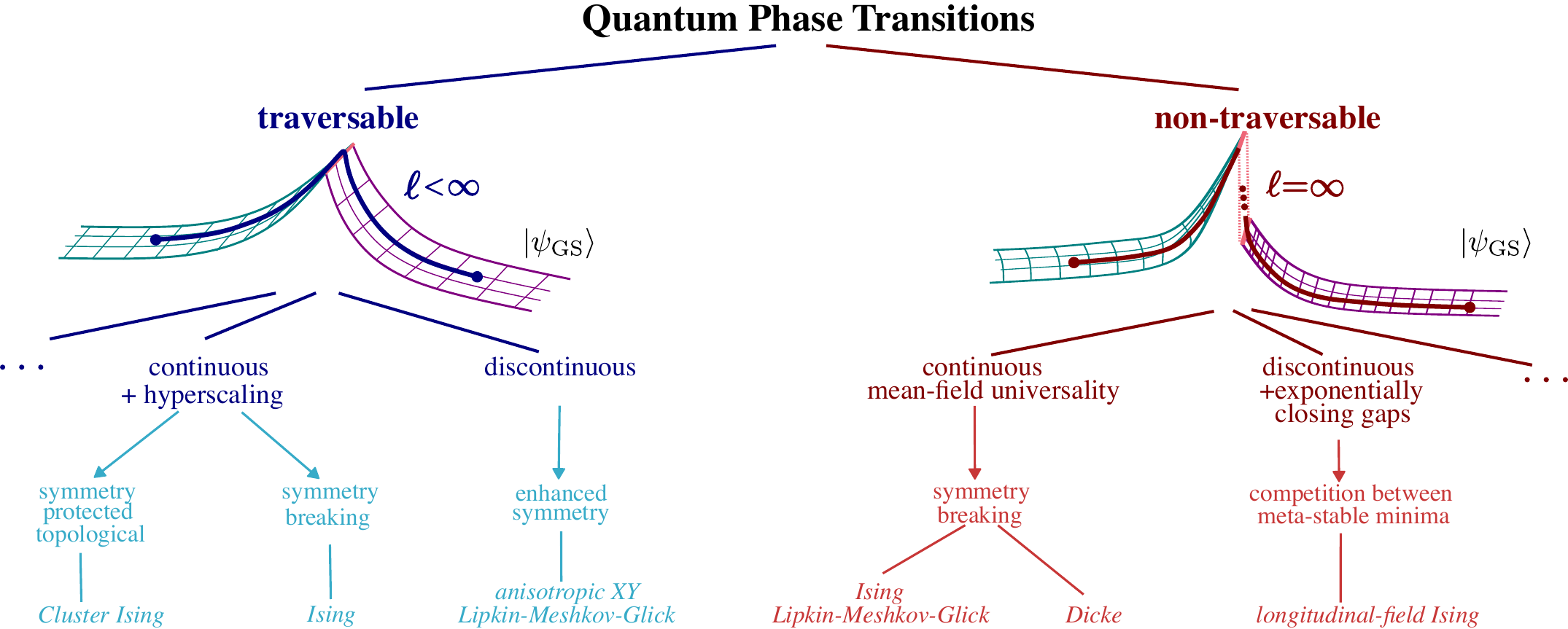}
    \caption{
        \textbf{Genealogy of Traversable Quantum Phase Transitions.}
        Quantum phase transitions can be divided into traversable (blue branch) and non-traversable (red branch) phase transitions according to the geometric length across the phase transition point remaining finite or not.
        Traversable quantum phase transitions, Sec.~\ref{sec:traversable}, include continuous ones obeying hyperscaling---including symmetry breaking phases (e.g., Ising chains, Sec.~\ref{sec:traversable:ssb}) and symmetry protected topological phases (e.g., Cluster Ising chain, Sec.~\ref{sec:CIM})---and discontinuous ones with an enhanced continuous symmetry (e.g., the anisotropic XY Lipkin-Meshkov-Glick model, Sec.~\ref{sec:traversable_1st-order}).
        Likewise, non-traversable quantum phase transitions include continuous ones of the mean-field universality class (e.g., Dicke model, Sec.~\ref{sec:cont-nontrav}), and discontinuous ones with exponentially closing gaps (Sec.~\ref{sec:1st_order_exp-gap}).
    }
    \label{fig:QPT-tree}
\end{figure*}

\section{Traversability of quantum phase transitions} 
\label{sec:theory}

Consider a system described by a many-body Hamiltonian $H(\lambda)$ that depends on an externally tunable parameter $\lambda$, and assume a QPT occurs at some $\lambda_c$, separating two distinct phases of matter. 
One can ask a simple question: \textit{Does the ground-state wavefunction $\ket{\psi_\text{GS}[\lambda]}$ change continuously across the transition point $\lambda_c$?}
We will see that a careful analysis will lead us to define a new geometric property of quantum phase transitions---traversability.

What makes the answer elusive and challenging to ascertain is the thermodynamic limit---an essential prerequisite for the existence of phase transitions. 
To see why, notice first that expectation values of observables, such as the ground state energy, are extensive and therefore diverge as the (linear) system size $L$ increases. This is straightforwardly fixed by considering densities: e.g.,\ the ground-state energy density remains finite in the thermodynamic limit. 

Second, as pointed out by Anderson~\cite{Anderson1967Infrared,*Anderson1967Ground}, two many-body quantum states $\ket{\psi_\text{GS}[\lambda]}$ and $\ket{\psi_\text{GS}[\lambda+\dd\lambda]}$ are orthogonal, in the thermodynamic limit, already for an infinitesimal change $\dd\lambda$. Formally, for a $d$-dimensional system, this follows from the overlap
\begin{eqnarray*}
    |\langle \psi_\text{GS}[\lambda]|\psi_\text{GS}[\lambda + \dd\lambda]\rangle|^2 
    &=& 1 - \chi_\mathrm{F}(\lambda,L) \, \dd\lambda^2 {+}\mathcal{O}(\mathrm{d}\lambda^3) \nonumber\\
    &=& \mathrm e^{-\chi_\mathrm{F}(\lambda,L)\dd\lambda^2 + \mathcal{O}(\mathrm{d}\lambda^3)},
\end{eqnarray*}
where the fidelity susceptibility $\chi_\mathrm{F}$, which quantifies the infinitesimal change of the ground state under parameter change, is extensive, $\chi_\mathrm{F}(\lambda,L){\sim} L^d$; hence, it diverges in the thermodynamic limit. Known as the \textit{orthogonality catastrophe}, formally this phenomenon implies the breakdown of wavefunction continuity w.r.t.~the scalar product in Hilbert space. While one can reasonably expect continuity issues to be more severe at phase boundaries, Anderson's orthogonality theorem already applies deep within phases of matter.

On the other hand, from experiments, it is natural to expect that adiabatic time evolution, when it exists, should be continuous.
Indeed, the physical ground-state manifold concerns only normalized states and ignores their overall phases; it is part of the projective Hilbert space where distances are measured w.r.t.~the Fubini-Study metric, whose diagonal elements are given by the fidelity susceptibility $\chi_\mathrm{F}(\lambda,L)$~\cite{Provost1980Riemannian,Anandan1990Geometry,Chruscinski2004Geometric,Carollo2020Geometry}. We therefore define the notion of continuity for ground state wavefunctions using the Fubini-Study \textit{metric density},
\begin{equation}
\label{eq:FS_metric}
    g_{\lambda\lambda}=\lim_{L\to\infty} L^{-d}\chi_\mathrm{F}(\lambda,L)\,  
\end{equation}
which remains finite within a phase.

Hence, the question about the continuity of the ground-state wavefunction across a quantum phase transition naturally leads us to examine the properties of adiabatic evolution on the ground-state manifold.

\subsection{Adiabatic and counterdiabatic driving} 
\label{sec:CD}

Quantum adiabatic evolution is a dynamical process in which the parameter $\lambda(t)$ is scheduled to change in time $t{\in}[0,T]$ so slowly that the rate of creating excitations into nearby levels remains small compared to the size of the energy gap $\Delta(\lambda)$ between the ground and excited states~\cite{Avron1999Adiabatic,Teufel2003Adiabatic,Marzlin2004Inconsistency,Amin2009Consistency}. In particular, a necessary condition for adiabaticity is the existence of an instantaneous energy gap throughout the entire schedule $\lambda(t)$. Intuitively, adiabaticity relates the total duration $T$ to the minimal spectral gap along the path~\cite{Albash2018Adiabatic}. 

Yet, in the thermodynamic limit, the ground-state energy gap $\Delta(\lambda)$ closes at the phase transition point $\lambda_c{=}\lambda(t_c)$~\cite{Sachdev2011Quantum}. As a result, the ground state $\ket{\psi_\text{GS}[\lambda]}$ undergoes a nonanalytic change as a function of $\lambda$, echoing the question we posed about its continuity.
Physically, this gap closing manifests in the breakdown of adiabatic driving across the phase transition. Analyzing and quantifying the properties of the created nonadiabatic excitations is at the core of the Kibble-Zurek mechanism~\cite{Kibble1976Topology,*Kibble1980Some,Zurek1985Cosmological,Zurek2005Dynamics,delCampo2014Universality}. 

Recently, new ideas known as \textit{shortcuts to adiabaticity} were conceived to implement transitionless evolution of eigenstates~\cite{Chen2010Fast,GueryOdelin2019Shortcuts,Duncan2025Taming}. Among them, counterdiabatic (CD) driving~\cite{Demirplak2003Adiabatic,*Demirplak2005Assisted,*Demirplak2008Consistency,Berry2009Transitionless,delCampo2013Shortcuts,Jarzynski2013Generating,Sels2017Minimizing,Claeys2019Floquet,Takahashi2024Shortcuts,morawetz2025universal,finzgar2025counterdiabatic,ohga2025improving,banerjee2026partial} proposes to add additional control terms to the Hamiltonian $H(\lambda)$, to engineer exact adiabatic dynamics in a finite time. This is achieved by driving the system with a CD Hamiltonian
\begin{equation}
    \label{eq:CD}
    H_\mathrm{CD}(t) = H(\lambda(t)) + \dot{\lambda}(t) A(\lambda(t)),
\end{equation}
where $A(\lambda)$ is the adiabatic gauge potential (AGP) designed to cancel all nonadiabatic transitions between the instantaneous eigenstates of $H(\lambda)$~\cite{Kolodrubetz2017Geometry}. 
Formally, CD driving is constructed to speed up the evolution following any control schedule $\lambda(t)$ traversed at an arbitrary speed $\dot{\lambda}(t)$~\cite{rolandcerf2002}---a freedom that we shall exploit to answer the ground-state continuity question. 

Exact CD driving requires resolving all gaps between energy levels; for generic quantum many-body systems, this is infeasible since they close exponentially with the system size $L$, causing the full gauge potential $A(\lambda)$ to become a nonlocal operator (with a divergent norm density as $L{\to}\infty$)~\cite{Saberi2014Adiabatic,Kolodrubetz2017Geometry,Sels2017Minimizing}. 

However, for gapped systems, a weaker form of \textit{ground-state CD driving} survives the thermodynamic limit, using the gauge potential~\cite{Kato1950Adiabatic}
\begin{equation}
    \label{eq:AGP_GS}
    A_\text{GS}(\lambda) = -i \comm{\Pi_\text{GS}[\lambda]}{\partial_\lambda \Pi_\text{GS}[\lambda]},
\end{equation}
with $\Pi_\text{GS}[\lambda]$ the projector onto the ground-state manifold. Whereas the full gauge potential $A(\lambda)$ cancels all nonadiabatic excitations, its ground-state counterpart precludes population from leaking out of the ground-state manifold only. In addition, $A_\text{GS}(\lambda)$ has a finite Hilbert-Schmidt norm density in the thermodynamic limit, and the theorems of quasiadiabatic continuation guarantee the existence of a quasilocal representative with almost-exponentially suppressed tails in space~\cite{hastings2005Quasiadiabatic,osborne2007simulating,hastings2010locality,bachmann2012automorphic}, cf.~App.~\ref{app:Kato_vs_Hastings}.

In principle, CD driving alleviates the problems of adiabatic evolution in gapped systems and guarantees unit-fidelity ground-state evolution in finite time. 
In practice, the ground-state AGP density is generally composed of an infinite number of increasingly nonlocal terms; however, these are cut off exponentially with distance at a scale set by the (inverse) ground-state energy gap $\Delta(\lambda)$. 
Hence, at the phase transition point when the gap closes, the spatial decay is no longer guaranteed, and the AGP norm density blows up [cf.~Sec.~\ref{sec:TFIC}]. 
Therefore, this formal solution is of limited practical value when crossing QPTs.

The specific form of the nonlocal terms in the AGP $A_\text{GS}(\lambda_c)$ arising at phase transition points depends on the details of the system of interest. 
For continuous phase transitions, however, the divergence of its norm in the thermodynamic limit is universal, since its square is related to the fidelity susceptibility $\chi_\mathrm{F}$ via
\begin{equation}
\label{eq:AGP_norm_FS}
    \| A_\text{GS}(\lambda,L) \|^2 = \chi_\mathrm{F}(\lambda,L) .
\end{equation}
Due to the nonanalyticity of the ground state wavefunction at the critical point, the fidelity susceptibility density
\begin{equation}
\label{eq:scaling-FS}
    \lim_{L\to\infty}L^{-d}\chi_\mathrm{F}(\lambda, L) \sim  |\lambda-\lambda_c|^{-\alpha_\mathrm{F}}\, ,
\end{equation}
and, by Eq.~\eqref{eq:AGP_norm_FS}, also the AGP density diverges at the QPT, with a characteristic exponent $\alpha_\mathrm{F}{>}0$. 
This divergence is directly related to the loss of locality of the AGP at the transition point. 
Physically, it means that even if one were able to implement the nonlocal operators needed for CD driving (e.g.,\ on a classical or quantum computer), their couplings would still require infinitely strong control amplitudes as a result of the closing gap; this seemingly renders CD driving ill-posed in the thermodynamic limit. 
The latter has placed a formidable bottleneck on scaling up numerical studies of CD driving across phase transitions to large systems~\cite{Passarelli2020Counterdiabatic,Grabarits2026Fighting}. 

A key technical contribution of this work is to address the divergence, Eq.~\eqref{eq:scaling-FS}, of the CD term arising at QPT points. We will identify the precise conditions under which the divergence of the AGP can be tamed, thus obtaining a well-defined (albeit nonlocal) AGP at the QPT point in the thermodynamic limit.

\subsection{Dynamical probes of phase-transition signatures in the ground-state manifold} 
\label{sec:schedules}

Na\"ively, one might be tempted to conclude that the divergence of the fidelity susceptibility at the QPT point, $\chi_\mathrm{F}{\sim}|\lambda-\lambda_c|^{-\alpha_\mathrm{F}}$, prevents the ground state from changing continuously across the phase boundary. 

To examine the problem from a different angle, recall that $\chi_\mathrm{F}$ also defines the diagonal elements of the Fubini-Study metric restricted to the ground-state manifold. It follows that the metric density $g_{\lambda\lambda}$, Eq.~\eqref{eq:FS_metric}, is finite everywhere in the thermodynamic limit, except at the phase transition point $\lambda_c$ where it has a singularity in accord with Eq.~\eqref{eq:scaling-FS}. 
However, the components of the metric tensor are coordinate-chart-dependent. Hence, what appears to be a divergence at the phase transition point may be removable by a suitable coordinate transformation~\cite{Kolodrubetz2013Classifying}. 

For one-dimensional manifolds---i.e.,\ curves---changing coordinate charts corresponds to reparametrization. 
We distinguish between 
(i) \textit{regular} parametrizations that can be approximated by linear functions ${\sim}(\lambda-\lambda_c)$ in the vicinity of the phase transition point $\lambda_c$, and
(ii) \textit{singular} parametrizations, ${\sim}|\lambda-\lambda_c|^{p}\,\text{sgn}(\lambda-\lambda_c)$, that traverse the phase transition point with vanishing speed controlled by the exponent $p{>}1$~\cite{Sen2008Defect,Barankov2008Optimal}.

Notice that the question about continuity of the wavefunction across the QPT pertains to the quasistatic properties of the ground-state manifold, and is by itself unrelated to dynamics. 
Similarly, the analysis in QPTs usually concerns their quasistatic properties, too:
one solves for the ground state at each fixed value of $\lambda$, and determines the critical exponents from the static data. In the quasistatic setup, reparametrizations cannot change the physics; they are merely an illusion of the choice of coordinate chart. 
Indeed, regular parametrizations do not affect the analysis of QPTs because they are linear at the critical point, and the proportionality factor does not affect the critical scaling behavior. 

Singular parametrizations, however, mess up both the analytical properties of observables and the values of the critical exponents, and are in this sense unphysical. To see this, let $H(\lambda){=}H_0{+}\lambda H_1$ exhibit a continuous phase transition at $\lambda_c{=}0$, where the second derivative of the ground-state energy $\partial^2_\lambda\langle H\rangle|_{\lambda_c=0}$ diverges. If, instead, one used the singular parametrization $H(\mu){=}H_0{+}\mu^3 H_1$, then $\partial^2_\mu\langle H\rangle|_{\mu_c=0} = \lambda^{4/3}\partial^2_\lambda\langle H\rangle|_{\lambda_c=0}$ acquires an extra factor $\lambda^{4/3}$ which can shift the divergence to a higher-order derivative; it also changes the characteristic critical exponent governing the divergence of the second derivative of the energy. The physics should, however, be the same. The apparent paradox is resolved by noticing that singular parametrizations spoil the operator expansion close to the critical point, in which the control parameter always appears linearly: $H {=} H_c {+} (\lambda{-}\lambda_c) O {+} \mathcal{O}((\lambda{-}\lambda_c)^2)$~\cite{Cardy1996Scaling}. 
Thus, to figure out a physical parametrization, one can expand the Hamiltonian in terms of the spectral content of the renormalization group fixed point~\cite{fisher1998renormalization}, determine the relevant operator $O$, and its coefficient defines the sought-after parametrization. This procedure fixes the coordinate chart for the metric tensor and resolves the parametrization ambiguity.
For these reasons, phase transitions are classified as discontinuous/continuous rather than first/second order in the modern literature~\cite{goldenfeld2018lectures}, respectively.

When considering dynamical processes, the problem becomes much richer: first, note that the quasistatic picture breaks down since nonequilibrium dynamics can excite the system out of the ground-state manifold. Second, each parametrization is physical since it corresponds to some dynamics leading to observables that one can measure. We therefore change nomenclature from parametrization to \textit{schedule}, $\lambda(t)$.

Regular schedules lead to Kibble-Zurek physics. The breakdown of adiabaticity when driving across continuous phase transitions can indeed be described via the Kibble-Zurek mechanism~\cite{Kibble1976Topology,*Kibble1980Some,Zurek1985Cosmological,Zurek2005Dynamics,delCampo2014Universality}, which relates the number of excitations (topological defects) to the speed at which the QPT is crossed.
The universality of the QPT---a quasistatic property---manifests in the Kibble-Zurek scaling of defects. Dynamics provides the means to probe this universality. 

Recall that the adiabatic gauge potential blows up at the critical point, as a manifestation of the physical singularity in the ground-state manifold there. As a result, counterdiabatic driving following a regular schedule breaks down at the critical point when the thermodynamic limit is approached: although regular CD evolution is excitation-free for any finite system size, it is not well-posed in the thermodynamic limit due to the divergence of the AGP norm density, cf.~Eq.~\eqref{eq:scaling-FS}.

Singular schedules, on the other hand, can lead to well-defined counterdiabatic driving even in the thermodynamic limit: if they vanish at the QPT point faster than the AGP density diverges, the norm density of the counterdiabatic term, $L^{-d/2}\|\dot\lambda A_\text{GS}(\lambda)\|$, can remain finite. 
Exact, singular-schedule CD driving then eliminates all traces of Kibble-Zurek physics, and instead lets the dynamics through the phase transition point, connecting in time the two adjacent phases. 
The analytical properties of the ground-state energy following the drive are determined by the power $p$ with which the singular schedule vanishes at the QPT, and---importantly---they are measurable. However, the resulting physics does not probe the universality of the QPT but the geometry of the ground-state manifold, cf.~Sec.~\ref{sec:traversability}. 
Related to this observation, we find that features of quantum phase transitions, such as the existence of a local order parameter or symmetry breaking, do not determine when singular schedules exist.

In the following, we provide an explicit construction of singular CD schedules that remain finite in the thermodynamic limit, i.e.,\ do not feature diverging amplitude or frequency scales as $L{\to}\infty$.


\begin{figure*}[t!]
    \centering
    \includegraphics[width=\linewidth]{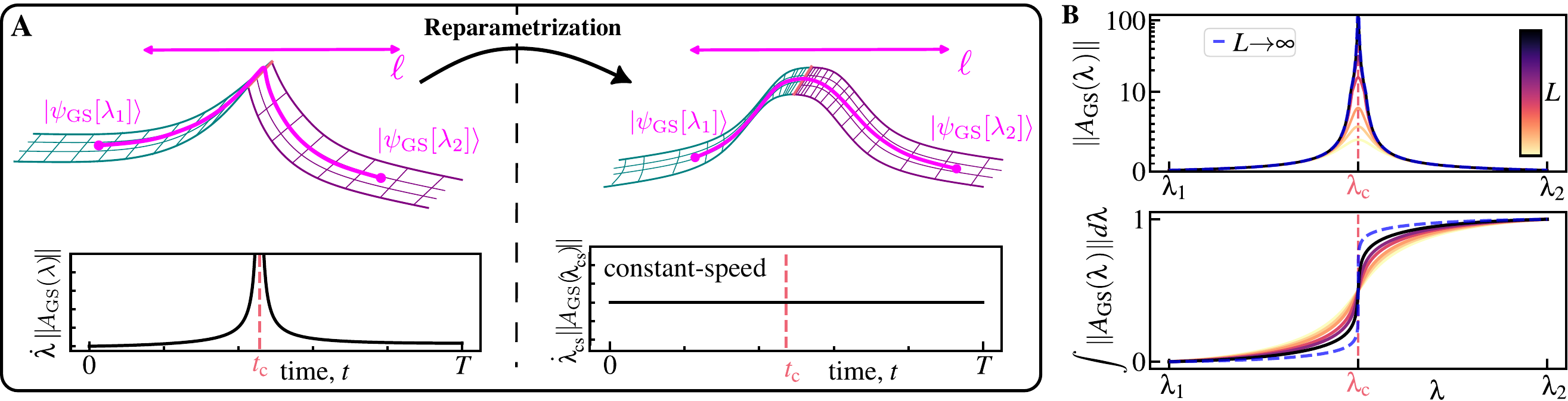}
    \caption{
        \textbf{Constant-speed counterdiabatic driving}. 
        \textbf{A}, geometric interpretation of constant-speed parametrization: using a singular reparametrization, the non-analyticity in the counterdiabatic term, $\dot\lambda A_\text{GS}(\lambda)$, at the quantum phase transition is cured.
        \textbf{B}, sketch of constant-speed reparametrization procedure:
        while the adiabatic gauge potential $\norm{A_\text{GS}}$ diverges, its integral across the quantum phase transition  $\int_{\lambda_1<\lambda_c}^{\lambda_2>\lambda_c}\norm{A_\text{GS}}\dd \lambda$ can remain finite, implying the existence of a finite-length path connecting the two ground states.
        The constant-speed parametrization follows this path at a constant speed $v$, i.e.,\ with each time-element $\dd t$ we increase $\norm{A_\text{GS}}\dd \lambda$ by a finite amount $v$.
        This cures the divergence at the quantum phase transition for $\dot{\lambda}A_\text{GS}(\lambda)$---the quantity entering the counterdiabatic driving term~\eqref{eq:CD}.
    }
    \label{fig:reparam}
\end{figure*}

\subsection{Constant-speed parametrization \& transcritical counterdiabatic evolution} 
\label{sec:reparametrization}

Observe that the ground-state AGP $A_\text{GS}(\lambda)$ always appears in the CD Hamiltonian in Eq.~\eqref{eq:CD} multiplied by the speed $\dot\lambda$.
It is then conceivable that singular schedules $\lambda(t)$ can exist, which slow down at the critical point faster than the AGP blows up. 
As a result, they lead to well-defined CD evolution in the TD limit, which eliminates excitations on top of the ground state, resulting in \textit{transcritical CD driving} sketched in Fig.~\ref{fig:reparam}.

To compensate for the divergence in $A_\text{GS}(\lambda)$ using $\dot{\lambda}$, we propose a \textit{constant-speed parametrization}, defined by\footnote{Another possibility would be to replace the $=$ sign with $\leq$, cf.~Eq.~\eqref{eq:AGP_norm_bound}; that would lead to a more complicated treatment without altering the conclusions significantly.}
\begin{equation}
\label{eq:constant-speed-condition}
    \dot{\lambda}(t) \lim_{L\to\infty} L^{-d/2} \norm{A_\text{GS}(\lambda(t),L)} = v,
\end{equation}
with $v$ an adjustable constant with units of energy density. 
This is a differential equation for the schedule $\lambda(t)$, supplemented with the endpoint (Dirichlet) boundary conditions $\lambda(0){=}\lambda_1$ and $\lambda(T) {=} \lambda_2$ set by the initial and target ground states.
Since the equation is first-order, the first boundary condition fixes the integration constant $t_0$, whereas the second fixes the value of $v$.\footnote{Alternatively, we can fix the available resources, $v$, and integrate from $\lambda_1$ to $\lambda_2$ within a time $T_v$ which is determined via integration.}
We denote the solution to Eq.~\eqref{eq:constant-speed-condition} by $\lambda_\mathrm{cs}(t)$, which is defined implicitly as the solution to the equation 
\begin{equation}
\label{eq:constant-speed-implicit}
    (t-t_0)v = \int_{\lambda_1}^{\lambda_\text{cs}(t)}\lim_{L\to\infty} L^{-d/2} \norm{A_\text{GS}(\lambda,L)} \dd \lambda .
\end{equation}
We provide examples for both analytical and numerical solutions in Secs.~\ref{sec:traversable} and~\ref{sec:nontraversable}.

Even though the norm density of the adiabatic gauge potential, $\lim_{L\to\infty} L^{-d/2}\norm{A_\text{GS}(\lambda,L)}{\sim} |\lambda{-}\lambda_c|^{-\alpha_\text{F}/2}$, diverges at the QPT point, the integral itself is finite for $\alpha_\mathrm{F}{<}2$. 
In such cases, Eq.~\eqref{eq:constant-speed-implicit} provides an explicit means to construct a singular schedule $\lambda_\text{cs}(t)$ that enables CD driving across the phase transition point with a fixed norm density of the counterterm $\dot\lambda A_\text{GS}(\lambda)$, set by the constant $v$. 
By design, the divergence of the AGP density is suppressed by the vanishing slope of the schedule at the QPT, $\dot{\lambda}_\text{cs}(t_c){=}0$, cf.~Fig.~\ref{fig:reparam}B.

Related ideas appeared in the context of adiabatic driving, where time reparametrization of the schedule was used to prove the efficiency of the adiabatic Grover algorithm~\cite{rolandcerf2002}. A large body of works ensued, leading to concepts like the quantum adiabatic brachistrochrone~\cite{Rezakhani2009Quantum} and the application of optimal control theory to modify annealing schedules~\cite{Grabarits2025Nonadiabatic}. In these works, however, the schedule $\lambda(t,L)$ explicitly depends on the system size $L$ since the adiabatic evolution is determined by the minimal gap $\Delta(L)$. As a consequence, the schedule $\lambda(t,L)$ becomes discontinuous in the thermodynamic limit if the total duration $T$ is kept finite~\cite{Grabarits2025Nonadiabatic}. 
By contrast, here we phrase the problem in terms of the physically meaningful densities and work directly in the thermodynamic limit; thus, we are only concerned with taming the divergence ${\sim}|\lambda-\lambda_c|^{-\alpha_\text{F}}$ while keeping the duration $T$ finite using CD driving. Crucially, all parameters defining the constant-speed schedule $\lambda_\text{cs}(t)$ (e.g.,\ amplitudes, frequencies, etc.) remain finite as $L{\to}\infty$. 

We emphasize that the constant-speed schedule does not eliminate the nonlocal terms arising in the adiabatic gauge potential at the phase transition point; moreover, these are essential to implement exact CD driving across the QPT. However, it suppresses their appearance in the CD Hamiltonian, placing a vanishingly small weight, $\dot\lambda$, up front, cf.~Eq.~\eqref{eq:CD}. Physically, this is possible because the nonlocality of the adiabatic gauge potential is isolated and only occurs at the QPT, whereas the AGP is a quasilocal operator at all other instances of the evolution. In Sec.~\ref{sec:meaning}, we show that this implies a finite norm density for the generator realizing the unitary rotation across the phase transition.

\subsection{Traversability criterion for quantum phase transitions} 
\label{sec:traversability}

Whenever they exist in the thermodynamic limit, constant-speed schedules provide an explicit means to evolve a ground state across a quantum phase transition using transcritical counterdiabatic driving.
This is enabled by identifying a singular parametrization of the path connecting the two ground states on either side of the transition, and may appear counterintuitive from the perspective of a diverging fidelity susceptibility density/metric tensor. 

For this reason, it was pointed out that geometric invariants (such as the Ricci scalar, or geodesic curvatures), rather than the components of the metric tensor $g_{\lambda\lambda}$, are the quantities to properly identify singularities in the ground state manifold, as they are parametrization-invariant~\cite{Kolodrubetz2013Classifying}. 
The only geometric invariant along the one-dimensional manifold defined by the control parameter $\lambda(t)$ is the distance  
\begin{equation}
\label{eq:def_ell}
    \ell = \int_0^T \dd t \dot \lambda \sqrt{g_{\lambda\lambda}} = \int_{\lambda_1}^{\lambda_2} \dd \lambda \sqrt{g_{\lambda\lambda}} .
\end{equation}
The line element $\sqrt{g_{\lambda\lambda}}\dd \lambda$ remains unaffected by reparametrizations of the manifold; hence, the distance $\ell$ can be written in the parametrization-invariant form above.

One now sees that the very existence of the constant-speed parametrization, Eqs.~\eqref{eq:constant-speed-condition}--\eqref{eq:constant-speed-implicit}, is inherently linked to a finite geometric distance between the initial $\ket{\psi_\text{GS}[\lambda_1]}$ and final $\ket{\psi_\text{GS}[\lambda_2]}$ ground states.
Indeed, using Eqs.~\eqref{eq:FS_metric} and~\eqref{eq:AGP_norm_FS} one can directly relate the metric tensor density to the norm density of the adiabatic gauge potential: 
\begin{equation}
\label{eq:intrinsic-length}
    \ell = \int_{\lambda_1}^{\lambda_2} \dd \lambda  \lim_{L\to\infty} L^{-d/2} \norm{A_\text{GS}(\lambda,L)}.
\end{equation}
Therefore, the constant-speed parametrization is an intrinsic-length parametrization, and the convergence of the integral in Eq.~\eqref{eq:constant-speed-implicit} for $\alpha_\mathrm{F}{<}2$ is equivalent to the existence of a finite geometric distance $\ell$.
It follows that the AGP norm density is integrable across the phase transition point only if the ground-state manifold is continuous in the thermodynamic limit; as a consequence, the ground-state wavefunction $\ket{\psi_\text{GS}[\lambda]}$ changes continuously across the QPT.

This prompts us to contemplate the possibility that some quantum phase transitions give rise to integrable singularities, and hence a finite-length distance $\ell$ on the ground state manifold between the two phases. One may prematurely attribute this feature to the nonlocality of the AGP density involved in the corresponding counterdiabatic evolution. 
Remarkably, we also find that certain commonly observed phase transitions do not admit a constant-speed CD evolution, and thus feature a divergent distance $\ell{\to}\infty$. 

We build on this observation to propose a dichotomous classification of QPTs according to whether the geometric distance $\ell$ across the transition point on the ground-state manifold is finite ($\alpha_\mathrm{F}{<}2$) or not ($\alpha_\mathrm{F}{\geq}2$). 
We term finite-length quantum phase transitions \textit{traversable}, since they can be dynamically driven through (at least in principle), using exact constant-speed counterdiabatic driving. All other quantum phase transitions we refer to as \textit{nontraversable}. 
In Sec.~\ref{sec:traversable}, we show that traversable QPTs include all critical points below the upper critical dimension as well as some discontinuous phase transitions with enhanced continuous symmetry, and provide concrete examples of unit-fidelity transcritical CD driving. By contrast, phase transitions in the mean-field universality class are nontraversable; we discuss these, as well as discontinuous phase transitions with competing metastable minima, in Sec.~\ref{sec:nontraversable}. 
Importantly, we find that traversability is a property unrelated to the existence of symmetry breaking, local order parameters, or universality; instead, it emerges as an intrinsic geometric feature of the ground-state manifold.

\begin{figure*}[t!]
    \centering
    \includegraphics[width=\linewidth]{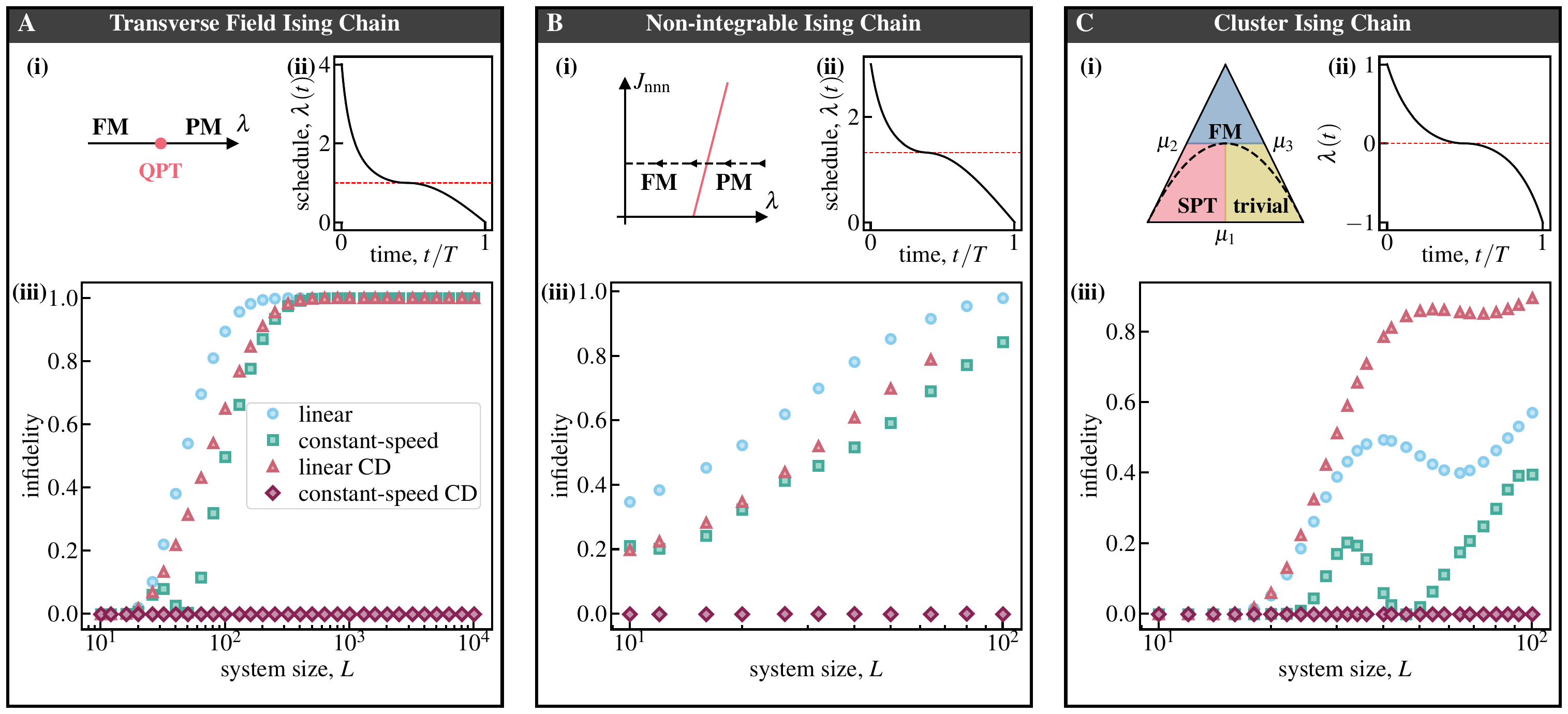}
    \caption{
    \textbf{Examples of traversable phase transitions} for transverse field Ising~\eqref{eq:H_TFIC} (A), and its non-integrable next-nearest neighbor version~\eqref{eq:TFIC_NNN} (B), and cluster Ising chain~\eqref{eq:H_CIM} (C); behavior for all three models is qualitatively same:
    \textbf{(i)} Sketch of traversed phase transition; dashed line indicates path through phase diagram.
    \textbf{(ii)} constant-speed schedules with characteristic slowing down at critical point; for TFIC and CIM one can obtain the exact $L=\infty$-schedule, for non-integrable Ising chain the schedule can be derived from finite system-size scaling, see App.~\ref{app:sec:numerics} for more details.
    \textbf{(iii)} Scaling of infidelity with system size at fixed ramp time $T$(${=}100,20,320$) for (A-C), respectively; while, finite norm liner ramp CD~(linear CD), the constant-speed and linear ramp schedules loose fidelity with increasing system size, the constant-speed CD schedule maintains unit fidelity up to the thermodynamic limit.
    The dips in the infidelity with system size for the cluster Ising chain (C) are likely related to the integrability of the model, causing spurious fidelity gains.
    Other parameters are: for the non-integrable Ising chain we choose $J_\text{nnn}=0.2$; for CIM the trajectory is described by $(\lambda_1, \lambda_2, \lambda_3)=(-(\lambda+1)^2,\ 2(\lambda-1)^2, \ (\lambda-1)^2)$ with $\lambda\in[-1,\,1]$; $\Omega$ for linear CD is determined by $\Omega=\norm{v}$ which is fixed by the constant-speed CD solution and ramp time $T$.
    }
    \label{fig:traversable}
\end{figure*}

\section{Operational measures of traversability} 
\label{sec:operational}

Formally, phase transitions occur in the strict thermodynamic limit, which does not exist in nature. This raises the question about defining operational measures of traversability that can be tested in experiments or numerical simulations with finite-size systems.   

Although the strict thermodynamic limit is most often inaccessible, its existence defines a \textit{thermodynamic regime} that allows us to probe the associated physics. Physically, the existence of a finite-speed parametrization translates to the existence of a finite-resource schedule whose amplitude and frequency scales remain bounded as the system size increases [Fig.~\ref{fig:reparam}B]. By contrast, some nontraversable QPTs, such as discontinuous phase transitions with exponentially closing gaps, require diverging amplitudes (i.e.,\ infinite-resource schedules) to connect the ground states on both sides of the transition, cf.~Sec.~\ref{sec:1st_order_exp-gap}. To reflect this limited energy budget, we set a uniform cutoff energy density $\Omega$ for the CD term density that is independent of the linear system size $L$ and the value of the parameter $\lambda$:
\begin{equation}
\label{eq:AGP_norm_bound}
    L^{-d/2}\| \dot{\lambda}(t) A_\text{GS}(\lambda(t)) \| \leq \Omega.
\end{equation}

In numerical simulations, imposing this energy-budget constraint allows us to systematically probe the traversability of QPTs in finite-size systems, as the system size $L$ increases. This enables us to compare the evolution of the ground state under the constant-speed parametrization with that under previously proposed schedules for unassisted and counterdiabatic evolution in Secs.~\ref{sec:traversable} and~\ref{sec:nontraversable}. 

In experiments, exact transcritical CD driving is mostly out of reach, since it requires the implementation of long-range multibody terms as a result of the closing energy gap in the vicinity of the phase transition point. Nevertheless, traversability constitutes an experimentally measurable property, as it is determined by the integrability of the square root fidelity susceptibility across the transition point, cf.~Sec.~\ref{sec:reparametrization}. 
The fidelity susceptibility is an experimentally accessible quantity, central to quantum metrology~\cite{rams2018atthelimits,zhang2026quantum}, that can be measured in linear response, e.g.,\ from the frequency-integrated dynamical structure factor~\cite{Kolodrubetz2017Geometry}, or the short-time limit of the Loschmidt echo following a small quench of a model parameter~\cite{gorin2006dynamics,karch2025probing}. For topological insulators in two dimensions, it is related to the Berry curvature~\cite{molignini2021crossdimensional} which is also experimentally measurable~\cite{Flaschner2016Experimental,Wimmer2017Experimental,Luu2018Measurement}. 
Moreover, some systems allow for reconstructing the divergence $\alpha_\text{F}$ of the fidelity susceptibility indirectly by measuring the scaling of alternative observables across the QPT using, e.g.,\ the Kibble-Zurek mechanism~\cite{delCampo2014Universality}. 

Notice that the constant-speed schedule can still be implemented even in situations where the nature of the phase transition is a priori unknown: since $\|A_\text{GS}(\lambda)\|$ can be 
evaluated at any fixed $\lambda$, the update rule
$
    \lambda_\text{cs}(t+\Delta t) = \lambda_\text{cs}(t) + {v\, \Delta t}/{\|A_\text{GS}(\lambda)\|}
$
can be stepped forward without prior knowledge of the transition's location or type---building the constant-speed schedule on-the-fly.
Traversability, and the corresponding constant-speed schedule, are then obtained from the finite-size scaling of the total schedule 
duration $T_\text{cs}(L)$: convergence signals traversability, divergence signals its absence. 
We detail this procedure in App.~\ref{app:sec:algorithm}.

To sum up, traversability is a physical property that can be determined both experimentally and numerically, and is related to the existence of a finite-resource schedule in the thermodynamic regime.

\section{Traversable quantum phase transitions} 
\label{sec:traversable}

In the following, we analyze specific models exhibiting QPTs, and demonstrate that (i) all continuous phase transitions that obey the hyperscaling relations~\cite{fisher1998renormalization} are traversable. 
In particular, we show this explicitly by exhibiting examples of both symmetry-breaking transitions, such as the paramagnet-to-ferromagnet (PM-FM) transition in Ising chains, and symmetry-protected topological (SPT)-to-trivial phase transitions in the cluster-Ising chain. 
We then show that (ii) traversability is not restricted to phase transitions with continuous order parameters, and discuss a traversable first-order transition in the anisotropic XY Lipkin-Meshkov-Glick model.

\subsection{Continuous transitions}

Continuous (second-order) phase transitions are characterized by an order parameter that vanishes continuously at the critical point. Systems undergoing continuous phase transitions exhibit critical phenomena; the transition points obey scale invariance, giving rise to universality that can be understood using renormalization group theory~\cite{fisher1998renormalization}. 

In the vicinity of the critical point, the fidelity susceptibility obeys a series of useful scaling relations, among which~\cite{venuti2007quantum,degrandi2010adiabatic,albuquerque2010quantum}
\begin{equation}
    \label{eq:scaling_FS}
    \chi_\mathrm{F}(\lambda, L) \sim L^d |\lambda-\lambda_c|^{-(2-\nu d)},
\end{equation}
which is valid iff hyperscaling~\cite{fisher1998renormalization} holds at the QPT. Here, $d$ is the spatial dimensionality of the system, and $\nu$ the correlation length exponent. As anticipated from Eq.~\eqref{eq:scaling-FS}, the fidelity susceptibility is extensive, and its density diverges as a power-law at the critical point. 
Scaling theory~\cite{fisher1998renormalization} allows us to reconcile the dependence on $L$ and $\lambda$ using a universal function $F(x)$,
\begin{equation}
\label{eq:scaling_F}
    \chi_F(\lambda,L) = L^{2/\nu}F\left( (\lambda-\lambda_c)L^{1/\nu}\right),
\end{equation}
where $F(x){\sim}|x|^{-(2-\nu d)}$ as $|x|\to\infty$, with $F(0)$ finite.

Since according to Eq.~\eqref{eq:def_ell} traversability is defined by a finite geometric length, using Eq.~\eqref{eq:scaling_FS}, one finds
\begin{eqnarray}
\label{eq:divergence_ell}
    \ell &{=}& \int \mathrm d\lambda \lim_{L\to\infty} L^{-d/2}\sqrt{\chi_F(\lambda,L)} \nonumber\\
    &{=}&
    \int_{\lambda_c-\epsilon}^{\lambda_c+\epsilon} |\lambda-\lambda_c|^{\nu d/2-1} \dd \lambda  + \text{regular} \ ,
\end{eqnarray}
where the geometric length was split into a regular and an irregular contribution.
The irregular contribution is finite, iff
\begin{equation}
\label{eq:traversability condition}
    \nu d >0,
\end{equation}
which is always realized since both $\nu{>}0$ and $d{>}0$.
Hence, one finds that all continuous phase transitions, which obey hyperscaling, are traversable. This class contains the large majority of critical phenomena. 
Therefore, we deduce that the ground state manifold for all systems obeying Eq.~\eqref{eq:scaling_FS} is continuous across the critical point. 

We emphasize that Eq.~\eqref{eq:scaling_FS} holds only under the assumption of hyperscaling, which is typically violated by long-range interacting systems or above the upper critical dimension, see Sec.~\ref{sec:nontraversable}.

To showcase traversability in practice, we now explicitly construct the constant-speed parametrization and use it to implement transcritical CD driving across the phase transition point in integrable and nonintegrable Ising chains. In the former case, we solve Eq.~\eqref{eq:constant-speed-implicit} analytically for the symmetry-breaking PM-to-FM transition in the transverse field Ising chain and the topological SPT-to-trivial transition in the cluster Ising chain; for the nonintegrable chain, we use tensor network techniques to determine the transcritical constant-speed schedule numerically. 

\subsubsection{Symmetry-breaking transitions}
\label{sec:traversable:ssb}

As toy models of a traversable symmetry-breaking transition, we use quantum Ising chains. First, we consider the integrable transverse-field Ising chain, where we demonstrate traversability by working out analytically the exact expression for the finite-speed schedule in the thermodynamic limit.  
Then, we add an integrability-breaking, next-nearest-neighbor interaction, which renders the model no longer exactly solvable. Using tensor-network simulations, we show how to compute the finite-speed schedule numerically via finite-size scaling and explicitly demonstrate traversability in the thermodynamic regime. Our transcritical CD driving approach to test traversability, therefore, applies to integrable and non-integrable systems alike.

\textbf{Integrable transverse-field Ising chain.} 
\label{sec:TFIC}
The Hamiltonian reads as
\begin{equation}
    \label{eq:H_TFIC}
    H(\lambda) = - \sum_{j=1}^L \left[\sigma_j^z \sigma_{j+1}^z + \lambda \sigma_j^x \right],
\end{equation}
where periodic boundary conditions are employed, and the overall energy scale is fixed to unity. The parameter $\lambda$ can be used to drive the system across a continuous QPT, separating a symmetric paramagnetic (PM) phase at $|\lambda| > 1$ from a symmetry-broken ferromagnetic (FM) phase at $|\lambda|<1$, see Fig.~\ref{fig:traversable}A(i).
We restrict the range to $\lambda {>}0$ by symmetry, and drive from the PM, $\lambda(0){=}\lambda_1{>}1$, to the perfect FM, $\lambda(T){=}\lambda_2{=}0$.

Using free-fermion techniques, the ground-state AGP can be computed exactly~\cite{delCampo2012Assisted}, cf.\ also Sec.~\ref{app:sec:TFIC}:
\begin{multline}
\label{eq:TFIC_AGP}
    A_\text{GS}(\lambda) = \frac{1}{8} \sum_{r \geq 1} \lambda^{r \mathrm{sgn}(1-|\lambda|)-1} \sum_j \left( \sigma_j^z \sigma_{j+1}^x \cdots \sigma_{j+r-1}^x \sigma_{j+r}^y \right. \\
    \left. + \sigma_j^y \sigma_{j+1}^x \cdots \sigma_{j+r-1}^x \sigma_{j+r}^z \right).
\end{multline}
The AGP is quasilocal away from the critical point: for $\lambda{\neq}1$, the coefficients decay exponentially in space. Exactly at the critical point, $\lambda_c{=}1$, the AGP becomes nonlocal with equal weight on operators acting at all distances.

The AGP norm density reads as (see Eq.~\eqref{app:eq:norm_AGP_TFIC})
\begin{equation}
\label{eq:norm_AGP_TFIC}
    L^{-1/2} \| A_\text{GS}(\lambda) \| =
    \begin{cases}
        [8(1-\lambda^2)]^{-1/2}     & \lambda<1 \\
        [8\lambda^2(\lambda^2-1)]^{-1/2}  & \lambda>1\,,
    \end{cases}
\end{equation}
respecting the universal power-law scaling, Eq.~\eqref{eq:scaling_FS}, with critical exponent $\nu\!=\!1$ and dimensionality $d\!=\!1$.
Crucially, since the singularity is integrable in the sense of Eq.~\eqref{eq:divergence_ell}, the QPT is traversable. 

Solving the differential equation for the constant-speed parametrization, Eq.~\eqref{eq:constant-speed-condition}, one obtains
\begin{equation}
\label{eq:cs_sol_TFIC}
    \lambda_\mathrm{cs}(t) = 
    \begin{cases}
        \displaystyle
        \frac{\tan[v(t-t_<)]}{\sqrt{1+\tan^2[v(t-t_<)]}} &\lambda\leq 1 \\[12pt]
        \displaystyle
        \sqrt{1+\tan^2[v(t-t_>)]} &\lambda\geq 1,
    \end{cases}
\end{equation}
where the constants $v,t_\lessgtr$ are fixed in terms of the boundary conditions, see Eqs.~\eqref{app:eq:TFIC_integration_constants}--\eqref{app:eq:TFIC_t_c}; the schedule $\lambda_\mathrm{cs}(t)$ is shown in Fig.~\ref{fig:traversable}A(ii). 
Importantly, as a direct consequence of traversability, all inherent energy scales of $\lambda_\mathrm{cs}(t)$, i.e.,\ $v,t_\lessgtr$, are intensive. 

We now benchmark the performance of transcritical CD driving following the constant-speed schedule $\lambda_\mathrm{cs}(t)$. First, we compare it against conventional CD driving with a linear-time schedule in the presence of equal energy budgets, cf.~Eq.~\eqref{eq:AGP_norm_bound},
where the constant $\Omega$ is fixed by Eq.~\eqref{eq:constant-speed-condition} as $\Omega = |v|$. 
When the energy-budget restriction is not satisfied, the AGP norm density is automatically capped at $\Omega$, reflecting the finite driving power available in experiments.

We do numerical simulations of the ground state evolution across the critical point with fixed increasing system sizes, and work in the thermodynamic regime identified in Sec.~\ref{sec:operational}.  
Figure~\ref{fig:traversable}A(iii) shows the infidelity as a function of system size following four different protocols: 
(1) The unassisted driving [blue circles], generated by $H(\lambda)$ with a linear schedule $\lambda(t)=\lambda_0(1-t/T)$, creates excitations across the critical point when the system size becomes large enough, as predicted by the Kibble-Zurek mechanism.
(2) Using a constant speed parametrization [green squares] slows down the drive at the critical point, which suppresses excitations intermittently, but the latter reappear as the system size increases.  
(3) CD driving using the linear schedule [red triangles] would normally suppress all excitations at any fixed system size; however, at larger system sizes, the divergent norm density of the GS AGP violates the energy budget restriction. Imposing the energy cap $\Omega$ leads to deviations from perfect CD evolution and hence to the build-up of infidelity.  
(4) By contrast, exact constant-speed CD driving [purple diamonds] leads to transitionless ground state driving across the critical point at a finite maximum energy cost $\Omega$. 

Whereas the continuity of the ground-state manifold is a quasistatic property ensured by the finite length connecting the ground states on both sides of the critical point, the protocol implementing transcritical driving, and hence traversability, is inherently dynamical. 
To see this, note that unitary dynamics is constrained by existing symmetries and the selection rules they impose.
Since the initial PM ground state at $\lambda_1{>}1$ is symmetric w.r.t.~the $\mathbb{Z}_2$ parity symmetry, transcritical CD driving preserves the parity sector. As a result, once the critical point is crossed, the system evolves into the cat-like superposition of the two degenerate ferromagnetic ground states. This requires the build-up of long-range entanglement, provided by the increasingly more nonlocal AGP close to the critical point, see Eq.~\eqref{eq:TFIC_AGP}. 

One may na\"ively fear that the exponentially-closing (with system size $L$) energy gap between the two FM ground states $\ket{\psi_{\text{GS}\pm}}$ will result in a rapidly vanishing energy denominator and hence give rise to an exponentially diverging AGP norm density already deep in the symmetry-broken phase; this is, however, not the case, since the gap denominator is suppressed by the vanishing matrix element $\bra{\psi_{\text{GS}-}}\partial_\lambda H\ket{\psi_{\text{GS}+}}{=}0$ in the numerator, cf.~Eq.~\eqref{eq:GS_Kato_AGP}.
We discuss this in the context of ramping the longitudinal field in Sec.~\ref{sec:1st_order_exp-gap}.

Conversely, notice that starting from the symmetry-broken FM ground state, the state will not evolve into the PM ground state because transcritical counterdiabatic driving is symmetry-preserving. Instead, the final state will be a controlled superposition of the ground and the lowest excited state. 
This counterdiabatic probe of traversability re-emphasizes the viewpoint that the latter is a dynamical rather than a purely quasistatic property.

\textbf{Nonintegrable Ising chain.} 
\label{sec:NITFIC}
While the integrability of the transverse-field Ising chain enabled the exact analytical derivation of the constant-speed schedule, for generic many-body systems, closed-form analytical solutions are unknown, and the transcritical CD schedule has to be constructed numerically. 

We illustrate the procedure to obtain the constant-speed schedule and confirm its superiority by simulating a non-integrable Ising chain with Hamiltonian\footnote{We also add a small boundary term $-0.1(\sigma_1^z+\sigma_L^z)$ to the Hamiltonian. This presents a common way to break the $\mathbb{Z}_2$ symmetry explicitly, and becomes irrelevant in the thermodynamic limit.}
\begin{equation}
\label{eq:TFIC_NNN}
	H(\lambda) = - \sum_j \left[ \sigma^z_j \sigma^z_{j+1} +  J_\text{nnn} \sigma^z_j \sigma^z_{j+2} + \lambda \sigma_j^x \right].
\end{equation}
We consider a next-nearest-neighbor interaction coupling $J_\text{nnn}{=}0.2$, which strengthens the ferromagnetic alignment and thus moves the QPT toward larger values of $\lambda$, i.e.,\ $\lambda_c{>}1$, see also Fig.~\ref{fig:traversable}B(i-ii).
The critical properties of the transition, however, remain unchanged, since the integrability-breaking perturbation $J_\text{nnn} \sigma^z_j \sigma^z_{j+2}$ is irrelevant in the renormalization-group sense; therefore, one expects the critical point to be traversable. Again, we drive from the PM, $\lambda(0){=}\lambda_1{>}\lambda_c$, to the perfect FM, $\lambda(T){=}\lambda_2{=}0$.

We implement the Hamiltonian via tensor networks~\cite{Fishman2022ITensor} and use the time-dependent variational principle~\cite{Haegeman2011Time} to implement the time evolution, keeping the bond dimension low enough for practical purposes; details are provided in App.~\ref{app:sec:NITFIC}. We stress that the usage of tensor networks is not strictly necessary, as any other way of implementing the time evolution, including the exact CD term, would yield the same results~\cite{Kim2024Variational,McKeever2024Towards,medvidovic2025adiabatic}. 

We present the numerical results in Fig.~\ref{fig:traversable}B.
As expected, constant-speed CD driving keeps unit fidelity with the target ground state across the phase transition for system sizes up to $L{=}100$; at the same time, the energy-capped linear-schedule CD creates noticeable excitations already at system sizes $L{\gtrsim} 10$. We emphasize that both CD protocols shown in Fig.~\ref{fig:traversable}B use the same numerically-determined ground-state AGP; they only differ in the driving schedule. Therefore, the failure of the linear-schedule CD driving at large system sizes demonstrates that access to the exact ground-state AGP, as provided, e.g.,\ by DMRG, is necessary but not sufficient for transitionless driving: the constant-speed schedule is the essential additional ingredient for traversability.

The simulations using the linear-schedule CD driving [red triangles] break down for $L{>}64$ due to uncontrolled volume-law entanglement growth caused by the accumulation of excitations when crossing the quantum critical point. By contrast, evolution following the constant-speed transcritical CD driving [purple diamonds] only needs to capture the critical logarithmic growth of entanglement, which is well within reach for the system sizes of interest in the thermodynamic regime (Sec.~\ref{sec:operational}). In particular, the infidelity remains zero within the tolerance of the simulation; this confirms that the truncation to a small operator bond dimension $\chi_\mathrm{AGP}=50$ captures the relevant physics.
We stress the crucial role of using the ground-state Kato AGP $A_\text{GS}$ for transcritical CD driving: the exponentially diverging norm density of the full AGP $A$ would not allow for an efficient compression of its bond dimension to produce an operational dynamical protocol, cf.~App.~\ref{app:Kato_vs_Hastings}. 

We thus see that transcritical CD driving using the constant-speed schedule is not only applicable in nonintegrable systems; it moreover requires fewer computational resources compared to conventional CD driving, and allows one to reach system sizes within the thermodynamic regime that are compatible with contemporary quantum simulators.

\begin{figure*}[t!]
    \centering
    \includegraphics[width=\linewidth]{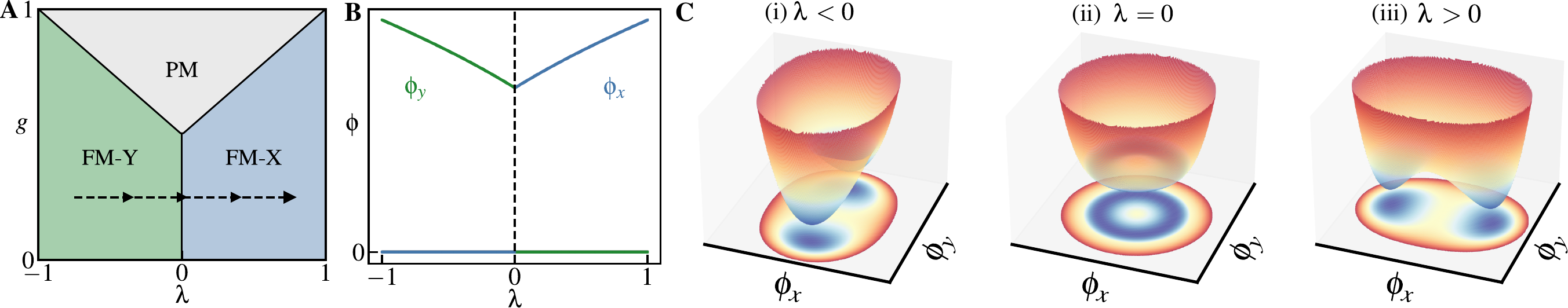}
    \caption{
    \textbf{Traversable discontinuous phase transition:}
    \textbf{A}, sketch of phase diagram of the anisotropic XY LMG model~\eqref{eq:LMG_H-1st_order}: at $g=g_c$ a continuous phase transition between a ferromagnet~(FM) and paramagnet~(PM) occurs (see also Fig.~\ref{fig:LMG_RCprotocl}); for $g<g_c$ a discontinuous phase transition between an X-ferromagnet and a Y-ferromagnet occurs at $\lambda{=}0$.
    \textbf{B}, order parameters $\phi_x$ and $\phi_y$ vs anisotropy parameter $\lambda$: at the phase transition, $\phi_{x}$ and $\phi_y$ undergo a discontinuous jump indicating the transition is discontinuous (first-order).
    \textbf{C}, (i) for $\lambda<0$, the Landau-Ginburg potential Eq.~\eqref{eq:lmg:landau-ginzburg} has two degenerate minima on the y-axis (FM-Y); (ii) at $\lambda=0$, the model becomes $U(1)$-symmetric and the groundstate manifold becomes a circle with no preferred direction; (iii) for $\lambda>0$, again two discrete minima along the $x$-axis (FM-X) occur. This enhanced symmetry enables a continuous deformation of the ground state between the two phases, and the phase transition is traversable. 
    }
    \label{fig:mexican_hat}
\end{figure*}

\subsubsection{Symmetry-protected topological transitions} 
\label{sec:CIM}

Symmetry-breaking transitions are characterized by a local order parameter, and one may wonder if this property is related to traversability. We therefore now briefly turn our attention to topological QPTs. 

The \textbf{Cluster Ising chain} Hamiltonian~\cite{Wolf2006quantum,Son2011Quantum,smacchia2011statistical,Verresen2017One,verresen2018edgemodes,ding2019phase,Smith2022Crossing}
\begin{equation}
    \label{eq:H_CIM}
	H = \sum_j \left[ \mu_1 \sigma_j^x + \mu_2 \sigma^z_j \sigma^z_{j+1} + \mu_3 \sigma^z_j \sigma^x_{j+1} \sigma^z_{j+2}\right],
\end{equation}
has an external $x$-field of strength $\mu_1$, a nearest-neighbor $zz$-interaction with coupling $\mu_2$ and a three-spin term of strength $\mu_3$. The ground state hosts three phases (see Fig.~\ref{fig:traversable}C(i)): a trivial one, a ferromagnetic one, and a symmetry-protected topological (SPT) one, depending on the values of $\mu_{1,2,3}$. The symmetry protecting the topologically ordered phase, which extends in the vicinity of the point $\mu_1 {=} \mu_2 {=} 0$, is $\mathbb{Z}_2 \times \mathbb{Z}_2$ generated by the global spin flip operator $\prod_i \sigma_i^x$ and time reversal (complex conjugation)~\cite{Verresen2017One,Smith2022Crossing}. 

Following Ref.~\cite{Smith2022Crossing}, we consider the path in parameter space
\begin{equation}
    \label{eq:CIM_path}
    \mu_1(\lambda) {=} {-}(\lambda{+}1)^2, \ \ \
    \mu_2(\lambda) {=} 2(\lambda^2{-}1), \ \ \
    \mu_3(\lambda) {=} (\lambda{-}1)^2,
\end{equation}
with $\lambda{>}0$ corresponding to the (trivial) product state phase, and $\lambda{<}0$ to the SPT cluster phase, characterized by a nonlocal string-order parameter; the two phases are separated by a topological critical point at $\lambda_c{=}0$ where the ground-state energy gap closes. We start from the ground state at $\lambda(0){=}\lambda_1{=}{+}1$ and drive the system to $\lambda(T){=}\lambda_2{=}{-}1$ through the topological phase transition. 

In the thermodynamic limit, the ground-state AGP has norm density [see App.~\ref{app:sec:CIM}]
\begin{equation}
    L^{-1/2}\| A_\text{GS}(\lambda) \| = \frac{1}{\sqrt{2 |\lambda|} (1+|\lambda|)}.
\end{equation}
The divergence at the phase transition point $\lambda_c{=}0$ leads to a finite integral in Eq.~\eqref{eq:constant-speed-implicit} and hence a finite geometric length. Therefore, the ground state manifold is continuous, and the SPT-to-trivial phase transition is traversable. 
Solving the differential equation~\eqref{eq:constant-speed-condition} analytically, one finds a particularly simple constant-speed schedule
\begin{equation}
    \lambda_\mathrm{cs}(t) = \pm \tan^2[v(t-t_0)].
\end{equation}
The parameters $v$ and $t_0$ are determined by the initial and final values $\lambda_{1,2}$, and the sign is chosen according to the region $\lambda{>}0$ or $\lambda{<}0$, respectively. The schedule, shown in Fig.~\ref{fig:traversable}C(ii), is finite in the thermodynamic limit. 

The numerical results in Fig.~\ref{fig:traversable}C confirm that the constant-speed CD protocol maintains unit fidelity across the SPT transition in the thermodynamic regime [cf.~Sec.~\ref{sec:operational}]. Similar to the Ising chains discussed above, it outperforms the alternative driving protocols while complying with a finite driving energy budget. 

We conclude that traversability can also apply to topological phase transitions and is, therefore, not related to the existence of a local order parameter.


\subsection{Discontinuous transitions}
\label{sec:traversable_1st-order}

One might suspect that traversability is correlated with the existence of a continuously vanishing (albeit possibly nonlocal) order parameter. We now move away from the realm of critical phenomena and present a model that exhibits a traversable discontinuous phase transition. 

The \textbf{anisotropic XY Lipkin-Meshkov-Glick model} is defined as
\begin{equation}
\label{eq:LMG_H-1st_order}
    H(\lambda) = -\frac{2}{L}\left(\frac{1+\lambda}{2}S_x^2+\frac{1-\lambda}{2} S_y^2\right)-2g S_z,
\end{equation}
where $S_{a} = \sum_{i = 1}^L \sigma_i^a$ represents the total spin operator along direction $a=x,y,z$ for an system composed of of $L$ spins-1/2, and we assume $\lambda\in[-1,1]$. This model exhibits a continuous, symmetry-breaking QPT at $g_c=(1+\abs{\lambda})/2$ between a doubly-degenerate FM at $g{<}g_c$ and a unique PM at $g{>}g_c$: we will analyze this continuous phase transition in detail in Sec.~\ref{sec:LMG-nontrav} (as it turns out to be nontraversable), see also Fig.~\ref{fig:mexican_hat}A.
Here, instead, we focus on the discontinuous transition which takes place at $\lambda_c {=} 0$ \emph{within} the FM phase $g_c{<}1$: the discontinuous transition separates a FM ground state polarized in the $x$ direction ($\lambda{>}0$) from one polarized in the $y$ direction ($\lambda{<}0$), see Fig.~\ref{fig:mexican_hat}B. Therefore, in this section, we fix the transverse field to be $|g|{<}1$ and consider the $xy$-anisotropy to be our time-dependent parameter $\lambda(t)$. 

In the thermodynamic limit $L\to\infty$, the ground-state sector is described by free bosonic excitations, corresponding to Gaussian quantum fluctuations around a mean-field spin-coherent state~\cite{dusuel2005continuous, defenu2018dynamical}.
A spin-to-boson mapping enables the analytical calculation of the fidelity susceptibility with respect to variations of the anisotropy parameter $\lambda$ and for fixed transverse-field $g<(1+\abs{\lambda})/2$ on both sides of the discontinuous transition (see App.~\ref{app:sec:LMG}). This leads to
\begin{align}
    \label{eq:FS_LMG-1st_order}\chi_F(\lambda|g<(1\pm\lambda)/2) = \frac{L}{2}\frac{g^2}{\abs{\lambda}(1+\abs{\lambda})^3}+\mathcal{O}(L^0)\,.
\end{align}
Then, in the thermodynamic limit, the ground-state AGP has norm density
\begin{align}
    L^{-1/2}\| A_\text{GS}(\lambda) \| = \frac{|g|}{\sqrt{2|\lambda|} (1+|\lambda|)^{3/2}}\,.
\end{align}
Remarkably, although the transition is discontinuous according to the conventional classification based on the behavior of the order parameter, the ground-state AGP norm density and correlation length\footnote{The notion of a correlation length can be extended to all-to-all coupled systems, see Eq.~\eqref{eq:corr-length_LMG} and the corresponding discussion.} diverge at the critical point $\lambda_c {=} 0$ as ${\sim} |\lambda|^{-1/2}$, so that the condition for traversability is satisfied. 

Indeed, solving the differential equation~\eqref{eq:constant-speed-condition} for the constant-speed parametrization, one obtains
\begin{align}
    \lambda_\mathrm{cs}(t) = \frac{(c_0-vt)^2}{(c_0-vt)^2-2g^2}\, \text{sgn}(t-t_c),
\end{align}
where the constant $c_0 = |g|\sqrt{2\lambda_1/(\lambda_1+1)}$ is fixed by the initial condition $\lambda_\mathrm{cs}(0) = \lambda_1>0$, and $t_c = c_0/v$ is the time at which the phase transition point $\lambda = 0$ is crossed.

The traversability of the discontinuous transition between the $x$-ferromagnetic and the $y$-ferromagnetic states can be understood via a Mexican-hat potential picture, as shown in Fig.~\ref{fig:mexican_hat}C. The idea is to consider the Landau-Ginzburg description of the anisotropic XY LMG model: the order parameter is the two-dimensional vector $\phi =(\phi_x,\phi_y)$ and the Lagrangian is
\begin{equation}
\label{eq:lmg:landau-ginzburg}
    L(\lambda) = \frac{1}{2} (\partial_\mu \phi_\mu)^2 +\left(\frac{1+\lambda}{2} \phi_x^2+\frac{1-\lambda}{2} \phi_y^2\right) - \frac{u}{4} |\phi|^4,
\end{equation}
where in the following we set $u{\equiv}1$ for simplicity.
For all $\lambda{>}0$, the potential looks as in Fig.~\ref{fig:mexican_hat}C(iii): the minima are at $\phi_{>,\pm}{=} (\phi_x,\phi_y){=}(\pm \sqrt{1+\lambda}, 0)$, and indeed the ground state is a FM in the $x$ direction. For every $\lambda{<}0$, instead, the minima are at $\phi_{<,\pm} {=} (0, \pm \sqrt{1-\lambda})$ and the ground state is a FM in the $y$ direction, see Fig.~\ref{fig:mexican_hat}C(i).
The two different ground states retain their identity until $\lambda$ crosses 0, where the potential acquires an enhanced U(1) symmetry (rotations in the $xy$-plane, see Fig.~\ref{fig:mexican_hat}C(ii)): only there does the ground-state manifold become a circle, the vacuum can acquire a nonzero expectation value in any direction, and a Goldstone mode signals the breaking of a continuous symmetry. 

With this picture in mind, the dynamical process of adiabatically (or counterdiabatically) passing from $\lambda{>}0$ to $\lambda{<}0$, and vice versa, goes as follows. For $\lambda{>}0$, start from the ground state formed by the equal weight, coherent superposition of $\ket*{\phi_{>,+}}$ and $\ket*{\phi_{>,-}}$, call it $\ket{\psi_>[\lambda]}$.
This parity-symmetric ground state is the actual ground state at any finite system size, whereas the parity-antisymmetric one is slightly more energetic. The ground state $\ket{\psi_>[\lambda]}$ can be expanded in eigenfunctions of rotations, e.g.\ by expressing in radial coordinates $(\phi_x,\phi_y) \longrightarrow (\phi_r,\phi_\theta)$ and then using the Fourier components $\phi_\theta {\sim} e^{il\theta}$. Because of the parity symmetry, there will be an $l{=}0$ contribution: $\ket{\psi_>[\lambda]} = c_0 \ket*{l=0} + \cdots$. The same happens at $\lambda{<}0$, where the corresponding $\ket{\psi_<[\lambda]}$ is the equal-weight superposition of the two degenerate ground states in the $y$ direction. The only missing point is $\lambda{=}0$, where the ground state (at any finite system size) is composed of \emph{only} the $l{=}0$ harmonic, all the other higher harmonics giving rise to a ground-state band. In conclusion, one can evince from this reasoning that there is room for a \emph{continuous} deformation of the ground state at $\lambda{>}0$ to the one at $\lambda{<}0$, provided one restricts to the parity-symmetric sector (which counterdiabatic evolution obeys). 

We stress again that traversability can be achieved only by restricting to the parity symmetric sector at each finite system size, and then taking the thermodynamic limit within the sector. The emergent degeneracy of all ground states of the form $\propto \alpha \ket*{\phi_{\lessgtr,+}} + \beta \ket*{\phi_{\lessgtr,-}}$ in the thermodynamic limit prevents traversability from one choice of $\alpha$,$\beta$ at $\lambda{>}0$ to another at $\lambda{<}0$. In a way, this is akin to the adiabatic continuation of states degenerate in energy, and it does not constitute a problem from a dynamical perspective, since the symmetry is respected by the Hamiltonian. This example highlights once again how traversability is a property inherently dynamical in nature. 

The enhanced continuous symmetry at the phase transition point provides a concrete mechanism enabling traversability of discontinuous transitions. The existence of other such mechanisms remains an interesting open question.

\section{Nontraversable quantum phase transitions} 
\label{sec:nontraversable}

In Sec.~\ref{sec:traversable} we showed that all continuous critical points, which obey the hyperscaling relations~\cite{fisher1998renormalization}, are traversable. To look for nontraversable phase transitions, we therefore turn our attention to regimes beyond the validity of hyperscaling. 

Hyperscaling is not expected to hold for generic discontinuous (first-order) transitions and Kosterlitz-Thouless critical points [see Sec.~\ref{sec:discussion}] with essential singularities. Moreover, it can be violated for certain continuous (second-order) transitions described by scale-invariant RG fixed points~\cite{goldenfeld2018lectures}: (i) in the presence of dangerous irrelevant operators at the fixed point (e.g.,\ $\phi^4$-theory in $d{=}4$ dimensions), which generate nonanalytic (e.g.,\ logarithmic) corrections to the scaling relations; (ii) above the upper critical dimension, where mean-field theory applies and the critical exponents saturate at their mean-field values; or (iii) at fixed points exhibiting anisotropic scaling (e.g.,\ Lifshitz transitions and multicritical points), where different spatial directions scale with different exponents. 
In such cases, one has to investigate on a case-by-case basis whether the critical point is traversable.

Below, we first demonstrate that continuous symmetry-breaking transitions within the mean-field universality class are nontraversable [Sec.~\ref{sec:cont-nontrav}]; this includes frequently encountered models, such as the Lipkin-Meshkov-Glick model for spontaneous nuclear spin polarization and the Dicke model of superradiance. 
Then, we briefly discuss generic discontinuous QPTs with exponentially closing gaps, and show that these are also nontraversable [Sec.~\ref{sec:1st_order_exp-gap}]. 
In both cases, the constant-speed condition, Eq.~\eqref{eq:constant-speed-condition}, does not lead to a well-defined constant-speed reparametrization $\lambda_\mathrm{cs}(t)$. As a consequence, there is no finite geometric length that connects the two ground states on the opposite sides of the phase transition point, where the ground-state manifold exhibits a discontinuity. 
Importantly, it follows that traversability is a geometric property, different from the common paradigm of continuous/discontinuous phase transitions. 

\subsection{Continuous transitions in the mean-field universality class} 
\label{sec:cont-nontrav}

As a first example of nontraversable critical points, we consider systems with an infinite coordination number. This class of models includes systems with cavity-mediated all-to-all interactions, such as the Lipkin–Meshkov–Glick (LMG) model~\cite{glick1965validity} and the Dicke model~\cite{dicke1954coherence,hepp1973onthesuperradiant}, and more generically systems falling in the so-called strong-long-range class~\cite{defenu2023longrange}. These systems are characterized by microscopic components that interact via a two-body potential that decays as a power law of the distance, $V(r)\propto r^{-\alpha}$, with a decay exponent smaller than the spatial dimensionality, $\alpha<d$.

The models mentioned above are in a mean-field universality class\footnote{This is to be distinguished from the alternative meaning of mean-field as in, e.g.,\ models on the Bethe lattice. Indeed, the transverse-field Ising model on the Bethe lattice admits an exact solution using Belief Propagation (or its variants), where the critical point shifts with the coordination number. Whereas the critical exponent of the magnetization takes on the mean-field value $\beta{=}1/2$, the fidelity susceptibility diverges with exponent $\alpha_\text{F}{=}1$, and hence the critical point is traversable.}. 
Consequently, the hyperscaling relations underlying the analysis in Sec.~\ref{sec:CD} do not necessarily apply~\cite{botet1982size,floressola2015finitesize}.
Indeed, as a result of the infinite-range nature of the interactions, the concepts of `dimensionality' and `length' lose their standard meaning. To perform a finite-size scaling analysis at criticality, the correlation length $\xi$ is replaced by a dimension-independent quantity, the \textit{coherence number} of sites $N_c$, which follows the scaling law~\cite{botet1982size}
\begin{align}
    N_c\sim |\lambda-\lambda_c|^{-\nu^*}\,,
\end{align}
with exponent $\nu^*$ governing the relation between thermodynamic and finite-size scaling exponents.
For a generic observable with critical behavior
\begin{align}
    O\sim |\lambda-\lambda_c|^{x_O}\sim N_c^{-x_O/\nu^*},
\end{align}
its finite-size scaling at the critical point follows by taking a coherence number of the order of the total system size: defining $L$ via $N_c{\sim} L^d$, one gets to $O\sim L^{-d x_O/\nu^{*}}$.The generalized coherence exponent $\nu^{*}$ is estimated by assuming that the critical behavior of the infinitely coordinated model corresponds to that of a short-range interacting system at its upper critical dimension $d_\text{uc}$, where mean-field theory becomes exact \cite{botet1982size,solfanelli2024universality}.
The coherence number is then related to the correlation length of the corresponding short-range mean-field model through
\begin{align}
    N_c\sim\xi^{d_\text{uc}}\sim|\lambda-\lambda_c|^{-\nu_{\mathrm{mf}}d_\text{uc}}\,,
\label{eq:corr-length_LMG}
\end{align}
which yields $\nu^{*}{=}\nu_{\mathrm{mf}}d_\text{uc}$, where $\nu_{\mathrm{mf}}$ is the mean-field correlation-length exponent. 

Applying this generalized scaling theory to the finite-size scaling of the fidelity susceptibility at the critical point yields
\begin{align}
    \chi_F(\lambda=\lambda_c)\sim L^{2/\nu^*}\sim L^{2/(\nu_{\mathrm{mf}}d_\text{uc})}\,.\label{eq:FS nustar}
\end{align}
Additionally, for the derivation in Sec.~\ref{sec:CD}, it was assumed that $\chi_F$ scales as $L^d$ far from the critical point on both sides of the transition. For short-range interacting models, this follows from dimensional analysis.
In general, this is not guaranteed to hold: it might be violated in particular in models with strong long-range interactions \cite{kwok2008quantum,gu2009scaling,gu2010fidelity}.
In this case, one needs to introduce two additional scaling exponents, $d^\pm$, determining the scaling $\chi_F$ above/below the critical point
\begin{align}\label{eq:dpm}
    \chi_F(\lambda\neq \lambda_c)\sim L^{d^\pm} \, .
\end{align}
To match the finite size scalings in Eqs.~\eqref{eq:FS nustar} and~\eqref{eq:dpm}, $\chi_F$ must diverge at the critical point as
\begin{align}
    \chi_F(\lambda,L)\sim L^{d^\pm}|\lambda-\lambda_c|^{2-\nu^*d^{\pm}}.
\end{align}
The traversability condition for continuous phase transitions from Eq.~\eqref{eq:traversability condition} therefore generalizes to $\nu^*d^{\pm}>0$.

\begin{figure*}[t]
    \centering
    \includegraphics[width=\linewidth]{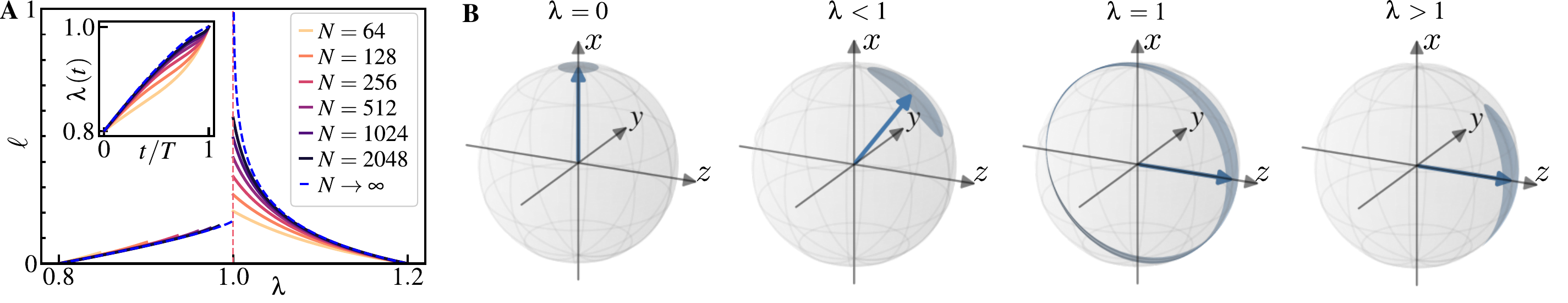}
    \caption{\textbf{Nontraversable continuous phase transition} in the Ising Lipkin-Meshkov-Glick model~\eqref{eq:LMG Hamiltonian}.
    \textbf{A}, geometric length $\ell$ as a function of final integration point $\lambda$ starting at initial point $\lambda_0$ in the ferromagnetic ($\lambda<1$, $\lambda_0=0.8$) and the paramagnetic ($\lambda>1$, $\lambda_0=1.2$) phase, for different system sizes $N=64,\dots,2048$ (yellow to black) and the thermodynamic limit result (blue), obrained by integrating Eq.~\eqref{eq:AGP norm density LMG FM} (for $\lambda<1$) and Eq.~\eqref{eq:FS LMG symmetric} (for $\lambda>1$).
    In the paramagnetic phase, the geometric length diverges at the quantum phase transition point ($\lambda=1$). 
    In contrast, geometric length remains finite in the ferromagnetic phase all the way to, and including, the quantum phase transition point; \textit{Inset} displays a corresponding constant-speed protocol enabling the preparation of a quantum critical state starting from the ferromagnet.
    \textbf{B}, sketch of continuous LMG ground state on the Bloch-sphere using the analytical expressions Eqs.~\eqref{app:eq:lmg_exp_fm} and~\eqref{app:eq:lmg_exp_pm}; blue arrow indicates the spin expectation value $\expval{S^{x,y,z}}$, and shaded region indicates spin-squeezing $\expval{\Delta S^{x,y,z}}$.
    In the ferromagnetic phase, the collective spin tips toward the $z$-axis with increasing $\lambda$, accompanied by spin squeezing in $y$-direction; in the paramagnetic phase, the collective spin expectation value remains pinned along $z$-direction and only fluctuations change with $\lambda$; fluctuations change discontinuously across the phase transition, and render it nontraversable.
    }
    \label{fig:LMG_RCprotocl}
\end{figure*}

\subsubsection{Lipkin-Meshkov-Glick model}
\label{sec:LMG-nontrav}

As a prototypical, infinitely correlated model we now consider the Lipkin-Meshkov-Glick (LMG) Hamiltonian
\begin{align}
\label{eq:LMG Hamiltonian}
    H(\lambda) = -\frac{2}{L}S_x^2-2\lambda S_z,
\end{align}
where $S_{a} = \sum_{i = 1}^L \sigma_i^a$ represents the total spin operator along direction $a=x,y,z$ for an system composed of of $L$ spins-1/2. Notice that we have already used an anisotropic version of this model in Sec.~\ref{sec:traversable_1st-order}, with a different naming of constants (to reserve $\lambda$ the parameter we drive). Recall that this model exhibits a symmetry-breaking QPT at $\lambda_c{=}1$ between a doubly-degenerate FM at $\lambda_c{<}1$ and a unique PM at $\lambda_c{>}1$. The upper critical dimension of the corresponding short-range model is $d_\text{uc}{=}3$ and its mean-field correlation-length exponent is $\nu_{\mathrm{mf}}{=}1/2$, yielding $\nu^*{=}3/2$. Therefore, the critical finite-size scaling is~\eqref{eq:FS nustar}
\begin{align}
    \chi_F(\lambda = 1)\sim L^{4/3},
\end{align}
in agreement with the numerical observations of Ref.~\cite{kwok2008quantum}.

In the thermodynamic limit $L\to\infty$, the ground-state sector is described by free bosonic excitations corresponding to Gaussian quantum fluctuations around a mean-field spin-coherent state~\cite{dusuel2005continuous, defenu2018dynamical}.
A spin-to-boson mapping enables the analytical computation of the fidelity susceptibility on both sides of the transition (see App.~\ref{app:sec:LMG}).
Since the critical exponents, $d^\pm$, on both sides of the critical point are not guaranteed to coincide, one needs to analyze them separately.

Let us start with the symmetry-broken ferromagnetic phase $0\leq\lambda\leq 1$, where
\begin{align}
    \label{eq:FS LMG paramagnetic}
    \chi^\text{FM}_\mathrm{F}(\lambda,L) = \frac{L}{4\sqrt{1-\lambda^2}}+\frac{\lambda^6}{32(1-\lambda)^2}+\mathcal{O}(L^{-1})\,.
\end{align}
Since the fidelity susceptibility is related to the adiabatic gauge potential [Eq.~\eqref{eq:AGP_norm_FS}], the latter follows the same structure: 
the term extensive in $L$ arises since the FM spin-coherent ground state changes direction as the parameter $\lambda$ is varied [Fig.~\ref{fig:LMG_RCprotocl}B], whereas the subleading correction serves to transport the quantum fluctuations. 

Deep inside the ferromagnetic phase ($0{<}\lambda{\ll}1$), $\chi_\mathrm{F}$ scales linearly with the system size $L$, giving $d^- {=} 1$; thus, the traversability condition $\nu^* d^- {= }3/2{>}0$ is satisfied.
In the thermodynamic limit, the ground-state AGP norm density reads as
\begin{align}\label{eq:AGP norm density LMG FM}
    \lim_{L\to\infty}L^{-1/2}\| A_\text{GS}^\text{FM}(\lambda,L) \| = \frac{1}{2(1-\lambda^2)^{1/4}}\,,    
\end{align}
and hence exhibits an integrable singularity as the critical point is approached, $\lambda{\to}\lambda_c^-{=}1$. Therefore, the critical point can be reached from the FM with finite geometric length, see Fig.~\ref{fig:LMG_RCprotocl}A. 
The corresponding constant-speed schedule is obtained by inverting
\begin{equation}
    F\big(\arcsin\lambda^\text{FM}_\text{cs}(t),2\big)-F\big(\arcsin\lambda_1,2\big) = vt\,,
\end{equation}
where $F(\phi,m)$ is the incomplete elliptic integral of the first kind~\cite{friedman1954handbook}, $\lambda_1$ and $v$ are set by the initial $\lambda_1=\lambda^\text{FM}_\text{cs}(0)$ and final $\lambda_2 =\lambda^\text{FM}_\text{cs}(T)$ conditions, and by the duration of the protocol $T$.
In Fig.~\ref{fig:LMG_RCprotocl}A, we compare the analytical solution in the thermodynamic limit (dashed blue line) to the geometric length and constant-speed schedule at finite system size $L$ (solid colored lines), obtained numerically by directly computing the exact ground state AGP with Eq.~\eqref{eq:AGP_GS}; the numerical results converge to the analytical solution in the thermodynamic limit.

By contrast, in the symmetric paramagnetic phase $\lambda{>}1$, one finds
\begin{align}\label{eq:FS LMG symmetric}
    \chi^\text{PM}_\mathrm{F}(\lambda,L) = \frac{1}{32\lambda^2(\lambda-1)^2}+\mathcal{O}(L^{-1})\,.
\end{align}
The absence of the extensive term is a consequence of the uniformity of the paramagnetic spin-coherent ground state, which remains pinned in the $z$-direction for all values of the parameter $\lambda$.
As a result, the ground-state AGP only has to transport the longitudinal Gaussian fluctuations, which are subextensive (see Fig.~\ref{fig:LMG_RCprotocl}B)---the relative fluctuations of the order parameter vanish as $\Delta S_{x/y}^2 {\sim} 1/\sqrt{L}$ in the PM phase. 

Deep in the paramagnetic phase ($\lambda{\gg}1$), the fidelity susceptibility is independent of system size, implying $d^+{=}0$ and thus violating the integrability condition~\eqref{eq:constant-speed-implicit}:
\begin{equation*}
    \lim_{L\to\infty} \int_{\lambda_c{+}\delta}^{\lambda_\text{cs}}\!\!\!\!\! L^{-\frac{d^+}{2}} \norm{A^\text{PM}_\text{GS}(\lambda,L)} \dd \lambda 
    \propto 
    \!\!
    \int_{\delta}^{\lambda_\text{cs}-1} \!\!\! \frac{1}{|\lambda|} \dd \lambda
    \propto \log\delta \overset{\delta\to 0}{\to} \infty .
\end{equation*}
Therefore the geometric length diverges in the paramagnetic phase when approaching the phase transition, see Fig.~\ref{fig:LMG_RCprotocl}A.
Consequently, reaching the critical point coming from the PM phase is impossible in the LMG model, even using counterdiabatic evolution. Geometrically, this means that there exists no finite-length path connecting the PM phase with the critical point, and hence the ground-state manifold is discontinuous.
Physically, approaching the phase transition point from the PM side requires a logarithmically divergent schedule time in the thermodynamic limit, and no regular constant-speed parametrization can remove this divergence. Curiously, transcritical CD driving does not fail by creating excitations; instead, its transitionless property forces the constant-speed schedule to slow down infinitely, and the state never reaches the critical point. 

We therefore conclude that the FM-to-PM continuous phase transition in the LMG model is nontraversable.
Physically, this nontraversability is deeply rooted in the structure of the ground state and the dominant role in the susceptibility played by quantum fluctuations in the PM phase:
in the thermodynamic limit, the LMG ground state is a Gaussian state, uniquely characterized by the first and second moments of the spin operators; a straightforward calculation exhibits the discontinuity of the relative fluctuations at the critical point, cf.~Eqs.~\eqref{app:eq:lmg_exp_fm} and~\eqref{app:eq:lmg_exp_pm}. 
Whereas the critical point can be reached in finite time starting from the FM side, there exists no constant-speed schedule that allows CD evolution to proceed and enter the PM phase.
Therefore, preparing the critical state proposed for sensing and metrological purposes~\cite{Degen2017Quantum,rams2018atthelimits,Liu2020Quantum} should be approached from the FM side.
Geometrically, this implies that the critical point belongs to the FM piece of the ground-state manifold.

\subsubsection{Dicke model of superradiance}
\label{sec:Dicke}

As a second example of a mean-field phase transition, we consider the Dicke model~\cite{dicke1954coherence}, which describes a system of $L$ two-level atoms interacting with a single monochromatic electromagnetic radiation cavity mode. The Hamiltonian reads as
\begin{align}
    H = \omega_ca^\dagger a +\omega_s S_z-\frac{\lambda}{\sqrt{L}}(a+a^\dagger)(S_++S_-),
\end{align}
where $\omega_c$ is the photon frequency, $\omega_s$ is the atomic energy levels separation, and $\lambda$ is the coupling strength. The atomic degrees of freedom are described by the total spin operators $S_{a} = \sum_{i = 1}^L \sigma_i^a$ along $a=x,y,z$, and the total spin raising and lowering operators $S_{\pm} = S_x\pm iS_y$. The photon's degrees of freedom correspond to the bosonic creation and annihilation operators $a^\dagger$ and $a$. 

Analogous behavior to the LMG case is expected, as both models belong to the same universality class. Indeed, both models are governed by collective infinite-range interactions and display mean-field–like critical behavior, with a critical mode described by an effective quadratic bosonic theory. This is supported by numerical estimates of the fidelity susceptibility scaling exponents on both sides of the transition, reported in Refs.~\cite{liu2005largeN,bastarrachea2014fidelity}. The fidelity susceptibility displays the same asymmetric finite-size scaling and identical critical divergences as in the LMG case, leading to the same breakdown of the integrability condition and, consequently, to the same kind of nontraversability of the critical point.

\subsection[Discontinuous quantum phase transitions with exponentially closing gaps]{Discontinuous quantum phase transitions \\ with exponentially closing gaps} 
\label{sec:1st_order_exp-gap}

We now investigate a qualitatively different mechanism for the breakdown of traversability in discontinuous (i.e.,\ first-order) QPTs with exponentially closing energy gaps. 

Since a comprehensive theory of discontinuous QPTs is still lacking, we refrain from formulating general conclusions, and instead provide an illustrative example using the mixed-field Ising chain
\begin{equation}
\label{eq:H_MFIM}
    H(\lambda) = - \sum_{j=1}^L \left[\sigma_j^z \sigma_{j+1}^z + g \sigma_j^x + \lambda \sigma_j^z \right],
\end{equation}
where we assume periodic boundary conditions. Notice that, unlike in Sec.~\ref{sec:TFIC}, here the drive parameter $\lambda$ controls the longitudinal field. In this model, one encounters a discontinuous QPT in the thermodynamic limit at $\lambda_c{=}0$, keeping $|g|<1$ fixed. Physically, it corresponds to the ground state switching from being a perturbatively-dressed ``all up'' state, denoted by $\ket{\Uparrow}$ ($\lambda{>}0$), to a perturbatively-dressed ``all down'' state $\ket{\Downarrow}$ ($\lambda{<}0$). 

Since the transition point is characterized by a finite correlation length, the low-energy sector consists of a nearly degenerate doublet, while the excitation gap to all other higher states remains finite at $\lambda_c{=}0$. 
We can therefore model the behavior of the system close to the transition using a simplified two-level system describing the two lowest-energy states~\cite{campostrini2014finite-size}:
\begin{eqnarray}
    H_\text{eff}(\lambda,L) \overset{|\lambda|\ll 1}{=} \varepsilon(\lambda,L)\tau^z + \Gamma(g,L) \tau^x, 
\end{eqnarray}
where $[\tau^\alpha,\tau^\beta]=2i\epsilon^{\alpha\beta\gamma}\tau^\gamma$ are effective Pauli matrices; the eigenstates of $\tau^z$ here correspond to the two finite-size dressed ferromagnetic states $\ket{\Uparrow}$ and $\ket{\Downarrow}$. The relative energy between these two states is $\varepsilon(\lambda){\sim} 2m_0 L \lambda$, and it arises from the energy cost due to the longitudinal term in Eq.~\eqref{eq:H_MFIM}, with ground-state magnetization density $m_0{=}(1-g^2)^{1/8}$ at $\lambda_c{=}0$~\cite{campostrini2014finite-size}. The tunneling matrix element $\Gamma$ is exponentially small in the system size $L$ at the transition point as a result of symmetry breaking; since the model is exactly solvable at $\lambda_c{=}0$, one is able to determine $\Gamma(g,L) {\approx} \sqrt{(1-g^2)/\pi L}g^L$, to leading order in $g$~\cite{Cabrera1987Role}.

The adiabatic gauge potential for the effective Hamiltonian $H_\text{eff}$ then reads as
\begin{eqnarray}
    A_\text{eff}(\lambda,L) = -\frac{1}{2}\frac{\Gamma(g,L)\partial_\lambda \varepsilon(\lambda,L)}{\varepsilon^2(\lambda,L)+\Gamma^2(g,L)} \; \tau^y.
\end{eqnarray}
The finite-size scaling regime describing the physics sufficiently close to the transition point is given by $L{\to}\infty$ and $\lambda{\to}0$, with the ratio $\varepsilon/\Gamma$ held fixed~\cite{rossini2016ground-state}.  
Then, the fidelity susceptibility diverges as 
\begin{equation}
\label{eq:suscept_1st-order}
    \chi_F(\lambda, L) = \frac{(\partial_\lambda \varepsilon)^2}{4\Gamma^2} \left(\frac{1}{1{+}(\varepsilon/\Gamma)^2}\right)^2
    \sim L^3 \mathrm g^{-2L}
    \sim \frac{L}{\lambda^2\log(1/|\lambda|)} ,
\end{equation}
where in the last step we used the scaling limit $\varepsilon/\Gamma=\text{const}$ to relate $L$ and $\lambda$ while keeping the fidelity susceptibility extensive.

The divergence of $L^{-1/2}\sqrt{\chi_F}$ close to the phase transition is so strong that it cannot be suppressed by any slowing-down of the schedule $\lambda(t)$, as can be verified explicitly using Eq.~\eqref{eq:constant-speed-implicit}.
Physically, this happens because, on either side of the transition, the two ground states are macroscopically distinct, and only an instantaneous macroscopic rotation can transform one into the other. To see this, notice that CD driving close to the transition is dominated by the heavy-norm AGP term. Consider the corresponding rotation $\exp(-i B_\text{eff})$, with
\begin{eqnarray}
    B_\text{eff} \!&\approx&\!\!\! \int \mathrm d\lambda A_\text{eff}(\lambda,L)
    = -\int \mathrm d\lambda \frac{\Gamma(g,L)\partial_\lambda \varepsilon(\lambda,L) }{\varepsilon^2(\lambda,L)+\Gamma^2(g,L)} \; \frac{\tau^y}{2} \nonumber\\
    &=& \!\!\! -\int \mathrm d\varepsilon \frac{\Gamma(g,L) }{\varepsilon^2+\Gamma^2(g,L)} \; \frac{\tau^y}{2} 
    \overset{L\to\infty}{\to} \int \mathrm d\varepsilon\; \pi \delta(\varepsilon) \;\frac{\tau^y}{2}   , 
\end{eqnarray}
where we used $\Gamma(g,L){\sim} \mathrm e^{-|\log g|L}{\to} 0$ independently of $\lambda$ as $L{\to}\infty$. Hence, the CD term in the thermodynamic limit implements an instantaneous $\pi$-rotation of the spins precisely at the transition point:
\begin{equation}
    \lim_{L\to\infty} \dot\lambda(t) A_\text{eff}(\lambda(t)) = \frac{\pi}{2}\delta(t)\, \tau_y \, ,
\end{equation}
which transfers the population between the two degenerate ground states. Notice that this delta-function pulse has a divergent strength, in stark contrast to the constant-speed schedule whose amplitude and frequency scales remain finite in the thermodynamic limit. This is a direct manifestation of the corresponding diverging geometric length across the transition point on the ground-state manifold which is, therefore, itself discontinuous. 

Finally, we emphasize the misleading simplicity of the AGP term, which appears to be generated by a single Pauli-$\tau^y$ matrix. In fact, the exponentially divergent norm in Eq.~\eqref{eq:suscept_1st-order}, and the dressing of the ferromagnetic states $\ket{\Uparrow}$ and $\ket{\Downarrow}$, imply that $A_\text{eff}(\lambda)$ grows nonlocal in the vicinity of the transition. It is an interesting open question whether the nonlocality at these first-order transitions is stronger or weaker compared to traversable QPTs, for which the gap closes polynomially.

In conclusion, discontinuous QPTs with an exponentially closing gap are nontraversable in a stronger sense than the continuous transitions in the LMG universality class. Whereas the latter admits reaching the critical point from the symmetry-broken phase but prohibits entering the symmetric phase, such discontinuous QPT points cannot be reached from either phase using a finite-time, finite-norm-density CD schedule. 
While our discussion was motivated by a specific example, the key ingredient of the analysis---the exponential closing of the energy gap with system size---is a feature shared by many different systems\footnote{There are known examples of first-order QPTs with polynomially closing gaps, see, e.g.,\ Sec.~\ref{sec:traversable_1st-order} or Ref.~\cite{Laumann2012Quantum}.}, such as ferromagnets, frustrated antiferromagnets, spin glasses, and some mean-field models~\cite{Amin2009First,Altshuler2010Anderson,Jorg2010Energy,Laumann2012Quantum}: all these are expected to exhibit strongly nontraversabe phase transition points.

\section{Some physical implications of (non-)traversability} 
\label{sec:meaning}

Let us now briefly discuss four physical implications of traversability, of both conceptual and practical significance.

\subsection{Geometry of the ground-state manifold}

As we alluded to in Sec.~\ref{sec:traversability}, one can understand traversability from quantum geometry. The family of ground states of a parameter-dependent Hamiltonian $H(\lambda)$ traces out a one-dimensional curve on the projective Hilbert space. At the QPT point, a nonanalyticity in this manifold develops in the thermodynamic limit, as indicated by the divergence in the Fubini-Study metric density, cf.~Eq.~\eqref{eq:FS_metric}. 
However, for traversable QPTs, the state manifold remains continuous; therefore, the nonanalyticity does not arise from a disconnected manifold, but from its curvature~\cite{Kolodrubetz2013Classifying}. 
By contrast, nontraversable QPTs require an infinite-norm (density) gauge potential at the transition point to connect the ground states across it, suggesting a direct discontinuity in the ground-state manifold in general\footnote{Mathematically, a curve may still be continuous even though its length diverges; we are not aware of a physical phase transition that exhibits this behavior.}.

\subsection{Quantum sensing \& metrology}

Since the fidelity susceptibility is directly proportional to the quantum Fisher information, the ground state continuity across the phase transition point has implications for quantum sensing and metrology~\cite{Degen2017Quantum,rams2018atthelimits,Liu2020Quantum}. The quantum Fisher information provides a limit on the precision with which a parameter can be estimated via the quantum Cram\'er–Rao bound~\cite{Helstrom1967Minimum,Degen2017Quantum}. Hence, states in the vicinity of phase transitions are often considered for parameter estimation and metrology due to their enhanced sensitivity to changes in the control parameter. 

Notice that reliable sensors require as well the stability of the parameter estimation procedure, which is related to the continuity of the ground-state manifold. 
For nontraversable phase transitions, the parameter window for precision estimation, set by the fidelity susceptibility density, shrinks with increasing system size; thus, this class of QPTs lacks robustness and provides only sharp threshold detection which may be difficult to implement in practice. By contrast, the continuity of the ground-state manifold in traversable phase transitions can be utilized for tunable sensing and precision estimation.
Thus, traversability implies that the enhanced sensitivity at the phase transition point is operationally accessible. 
As a perhaps counterintuitive consequence, for the Lipkin-Meshkov-Glick model, our work suggests that the traversable discontinuous phase transition between the ferromagnetic states (Sec.~\ref{sec:traversable_1st-order}) is better suited for quantum sensing than the nontraversable continuous transition between the ferromagnetic and paramagnetic states (Sec.~\ref{sec:LMG-nontrav}). 

\subsection{Unitary circuit complexity}

As a result of the ground-state continuity, the geometric length remains finite for traversable QPTs. In this case, one can convince oneself that the path-ordered unitary $\mathcal{P}\exp\left(-i\int^{\lambda_2}_{\lambda_1}\mathrm d\lambda A_\text{GS}(\lambda)\right){\equiv}\exp(-iB)$ connecting the ground states on the opposite side of the transition ($\lambda_1{\leq}\lambda_c{\leq}\lambda_2$), has an instantaneous generator $B$ of a finite Hilbert-Schmidt norm density, $\lim_{L\to\infty}L^{-d/2}\|B\|\leq\infty$.    
One may then na\"ively suppose ground states in traversable transitions to be easier to prepare on a quantum device than nontraversable states, e.g.,\ requiring a smaller circuit complexity; however, as we now elaborate, the nonlocal AGP still requires an $\mathcal{O}(L^d)$-depth local unitary circuit.

On the one hand, under the constant-speed parametrization, the CD term $\dot{\lambda}(t) A_\text{GS}(\lambda(t))$ generates a unitary evolution that maps $|\psi_\text{GS}[\lambda_1]\rangle$ to $|\psi_\text{GS}[\lambda_2]\rangle$ exactly while retaining a constant (and hence finite) norm density $\dot{\lambda} L^{-d/2} \|A_\text{GS}\|$, including at the critical point (where $\dot{\lambda} {\to} 0$ compensates for the divergence of $L^{-d/2}\|A_\text{GS}\|$).
When implemented on a digital quantum platform via Trotterization, such a generator gives rise to a circuit of depth proportional to the total evolution time $T$ times the norm density, divided into layers of bounded-norm gates~\cite{Lloyd1996universal,Childs2018toward}.
Therefore, it appears that the constant-speed parametrization induces a finite-depth unitary circuit connecting two phases of matter.

On the other hand, however, since the CD-assisted constant-speed parametrization reproduces exact adiabatic evolution, it is not clear how it should lead to an improved circuit complexity.
Indeed, as we discussed in Sec.~\ref{sec:CD}, the required generator $A_\text{GS}$ is highly nonlocal at the phase transition point $\lambda_c$. Hence, despite its finite norm density, the instantaneous generator $B$ is nonlocal too. This is easily seen in the transverse-field Ising chain, where the path-ordering in the definition of $B$ drops out. Integrating Eq.~\eqref{eq:TFIC_AGP} across the critical point then produces a spatial decay for the Pauli strings of $B$ of the form $1/r$, which keeps the Hilbert-Schmidt norm density of $B$ finite since $1/|r|^2$ is integrable in one dimension; coincidentally, the $1/r$ power-law decay also exhibits $B$ as a nonlocal operator, where $n$-body Pauli strings carry nonnegligible weight. 
Implementing such $n$-body terms with local two-qubit unitary gates generally requires an $\mathcal{O}(n)$-depth circuit~\cite{Nielsen_Chuang_2010,Whitfield2011simulation}.
Therefore, traversability does not compromise the definition of a quantum phase of matter~\cite{hastings2010locality}. 

Notably, access to such multi-body operations substantially reduces the circuit complexity, enabling a constant depth unitary that connects different phases of matter~\cite{moore1999quantumcircuitsfanoutparity,Hoyer2005fanout,Baumer2025Measurement}. In this context, the present study provides a concrete prescription for identifying the underlying unitary transformations in the case of traversable phase transitions.
It also suggests that finite-depth circuits containing long-range and multi-body operations may not be sufficient to adiabatically connect ground states of two phases of matter separated by a nontraversable phase transition, and exhibits concrete models for future study cases. 

\subsection{Adiabatic quantum computation}

From a theoretical perspective, our work imposes performance bounds on (CD-assisted) adiabatic quantum computing.
Specifically, note that, in the thermodynamic limit, nontraversable QPTs cannot be crossed using finite resources (time or amplitude), even when resorting to nonlocal Hamiltonians.
In Sec.~\ref{sec:nontraversable}, we showed that QPTs typically encountered in adiabatic quantum computing when solving NP-hard optimization problems---i.e.\ fully connected models and discontinuous phase transitions with an exponentially closing gap---fall precisely into the nontraversable category.

These observations provide a new geometric perspective on the anticipated difficulty of using counterdiabatic techniques for adiabatic quantum computation~\cite{Passarelli2020Counterdiabatic,Grabarits2026Fighting}: the transitions that encode computational hardness are precisely those that cannot be traversed with a finite-resource schedule.
Hence, traversability, in the context of adiabatic quantum computation, is a geometric signature of the absence of a computational barrier associated with the phase transition---a necessary (though not sufficient) condition for the QPT to be dynamically manageable.
We therefore conclude that typical NP-hard optimization problems involving nontraversable QPTs with exponentially closing gaps, cannot be solved efficiently using CD-assisted adiabatic quantum computation.

\section{Discussion \& Outlook} 
\label{sec:discussion}

The central result of this work is a universal dichotomous classification of quantum phase transition points as \textit{traversable} or \textit{nontraversable}. 

We showed that, by its origin, traversability can be traced back to the geometry of the ground-state manifold. 
In particular, traversable phase transitions exhibit a finite geometric distance across the transition point and, therefore, the ground-state manifold remains connected; the phase transition point itself introduces a curvature singularity, rather than an intrinsic divergence.
By contrast, for nontraversable QPTs, the ground state manifold itself is singular at the transition point, i.e.,\ its divergence cannot be removed by reparametrizing the coordinate chart.

Our analysis shows that all continuous phase transitions, which obey hyperscaling, are traversable (Sec.~\ref{sec:traversable}).
We have explicitly demonstrated this for both symmetry-breaking transitions, displayed by integrable and nonintegrable Ising chains, and the symmetry-protected topological transition in the cluster Ising chain. 
Furthermore, we have exhibited a generic mechanism by which discontinuous symmetry-breaking transitions with enhanced continuous symmetry at the transition point exhibit a continuous ground state manifold, and are hence also traversable; as a concrete example, we discussed the $x$-ferromagnet-to-$y$-ferromagnet transition in the anisotropic XY Lipkin-Meshkov-Glick (LMG) model.
By contrast, we have found that continuous symmetry-breaking transitions in strong-long-range and all-to-all interacting models, characteristic of the mean-field universality class above the upper critical dimension, are nontraversable (Sec.~\ref{sec:nontraversable}): examples include the continuous ferromagnet-to-paramagnet transition in the Ising LMG model and the superradiance transition in the Dicke model, where order parameter fluctuations play disproportionate roles for adiabatic transport in the two phases.
By the same mechanism, other strong-long-range interacting models are also expected to be non-traversable, like the Rabi-Stark model~\cite{chen2020quantum}---that is outside of the LMG/Dicke universality class but has the same $\nu^\ast d=0$ exponent.
Another generic class of nontraversable QPTs is discontinuous phase transitions with exponentially closing gaps.

These case studies clearly indicate that traversability is a classification of QPTs, distinct from the established continuous/discontinuous paradigm based on the continuity of the derivatives of the ground-state energy density. In particular, traversability is unrelated to symmetry breaking or the existence of local order parameters, and it is not limited to renormalization group fixed points. 

Beyond these classes, traversability may be examined on a case-by-case basis.
For instance, the Berezinskii-Kosterlitz-Thouless transition is characterized by a finite fidelity susceptibility~\cite{sun2015fidelity,cincio2019universal}; hence, it is most likely traversable, or at least has a reachable transition point when approached from the gapped phase, although additional analysis is required to confirm this indication.

It is a rather striking fact that the majority of the well-known phase transitions appear to be traversable, despite a divergent fidelity susceptibility.
Indeed, we exhibited only two general mechanisms for the breakdown of traversability; it is a challenging open problem to identify more. 
Interesting directions for future research include possible generalizations of the traversability concept to thermal phase transitions~\cite{basri2025thermodynamic}, disordered systems~\cite{Bray1980Replica,Igloi2005Strong,Evers2008Anderson,Abanin2019Colloquium}, dynamical~\cite{Heyl2018Dynamical} and excited-state/microcanonical~\cite{caprio2008excited,santos2016excited,zunkovic2018dynamical,cejnar2019static} QPTs, and phase transitions in nonequilibrium steady states~\cite{Hinrichsen2000Nonequilibrium}.
Nonequilibrium phase transitions occurring in periodically driven (Floquet) systems~\cite{schindler2024counterdiabatic,schidler2025geometric} are likely also amenable to the traversability classification, since one can define a proper eigenstate manifold whose geometry to analyze.

The quantum geometric interpretation of traversability promotes it to a dynamical concept. 
We demonstrated that traversable phase transition points can be crossed with unit fidelity using exact counterdiabatic (CD) driving in the thermodynamic limit. To do this, we identified the one-form $\dot\lambda A_\text{GS}(\lambda)$ defining the counterdiabatic term in the Hamiltonian as the physically relevant quantity, rather than the adiabatic gauge potential (AGP) alone.  
This insight allowed us to identify a singular finite-speed schedule that slows down at the phase transition point, $\dot\lambda {\to} 0$, faster than the AGP diverges $\norm{A_\text{GS}} {\to} \infty$, leading to well-defined transcritical CD driving in the thermodynamic limit with finite norm density at all times.  
Therefore, the divergence of the adiabatic gauge potential at a QPT does not, by itself, prevent exact counterdiabatic driving in the thermodynamic limit, nor does it indicate the breakdown of adiabatic evolution.

We have established a direct link between the existence of the finite-speed schedule and the traversability of the underlying QPT, and developed both analytical and numerical techniques to compute it in integrable and nonintegrable models. 
Importantly, the schedule has a well-defined functional form in the thermodynamic limit, i.e.,\ it does not exhibit divergent amplitude or frequency scales as the system size increases. 

Nevertheless, transcritical unit-fidelity CD driving requires an exact adiabatic gauge potential which becomes nonlocal at the phase transition point. Therefore, traversability complies with the definition of a quantum phase of matter.
We emphasize that traversability does not imply the existence of local dynamics through critical points; instead, the term derives from the finite-length path across the phase transition point within the ground-state manifold. 

The above considerations suggest that traversability is operationally dynamical. In particular, the corresponding counterdiabatic dynamics is symmetry-preserving and obeys any selection rules arising from symmetries; as a result, starting from a symmetry-broken state, it does not lead to the symmetric ground state on the other side of the transition. 
Nevertheless, unlike the Kibble-Zurek mechanism~\cite{Kibble1976Topology,*Kibble1980Some,Zurek1985Cosmological,Zurek2005Dynamics,delCampo2014Universality}, transcritical CD driving is by definition transitionless and hence cannot be used to probe the critical properties of phase transitions. 
We therefore conclude that, at its core, traversability is an inherently dynamical property, tightly linked to equilibrium critical behavior via the geometry of the ground-state manifold.

Even though the strict thermodynamic limit, in which phase transitions formally occur, is often inaccessible in numerical and experimental studies, we identified a thermodynamic regime that allows us to operationally access the relevant physics by extrapolating to larger system sizes. 
Physically, traversability quantifies whether a QPT can be crossed using finite-speed CD driving with finite resources; in practice, it defines a universal bound on how fast a control parameter can approach the phase transition point as the system size increases. 
As a result, we argued that traversability determines the operational suitability of phase transitions for quantum sensing. 
Moreover, it also implies a computational barrier for solving NP-hard optimization problems using adiabatic quantum computing, originating from phase transitions with exponentially closing gaps. 

In experiments, the traversability of a given phase transition can be operationally established by measuring the fidelity susceptibility via, e.g.,\ linear response. 
We note that the difficulty of dynamically implementing the transcritical counterdiabatic protocol scales with increasing the system size due to the nonlocality of the adiabatic gauge potential at the phase transition.
However, in this work, we focused on the unit-fidelity preparation of ground states; for approximate preparation of states in finite-size systems using ramps across critical points~\cite{Semeghini2021probing,Sun2023engineering,Lonard2023realization,karch2026dynamical,Evered2025probing,machado2026dipolar}, the constant-speed schedule using even approximate gauge potentials may already be beneficial, e.g.,\ for the preparation of quantum puddle/lake states~\cite{gjonbalaj2025shortcuts,karch2026dynamical}.
Moreover, the relevant thermodynamic regime is within reach in large digital quantum simulators that are developing a capability to implement nonlocal operators~\cite{Endres2011observation,Semeghini2021probing,pennington2026preparing}.
In addition, the ability to reduce the circuit complexity using adaptive measurement-and-feedforward circuits~\cite{Gottesman1999,Tantivasadakarn2023Hierarchy,Malz2024preparation,Iqbal2024Topological,Smith2024Constantdepth} may enable implementing nonlocal AGPs without requiring deep unitary circuits.
Therefore, the interplay between traversability and transcritical counterdiabatic driving can serve as a productive testing ground for the fundamental organizing principles of quantum matter and geometry, in the vicinity of quantum criticality.

\section*{Author Contributions} 

FB and PMS conceived the idea.
FB carried out the study of the Ising models.
AS performed the study of the Lipkin-Meshkov-Glick models.
MB supervised the project.
All authors contributed to writing the manuscript.

\acknowledgments

It is a pleasure to acknowledge discussions with Wojciech De Roeck, Beno\^it Dou\c{c}ot, Wen-Wei Ho, David A.\ Huse, Christopher Jarzynski, Hyeongjin Kim, Michael Kolodrubetz, Alessio Lerose, David M.~Long, and Anatoli Polkovnikov. 

This work was funded by the European Union (ERC, QuSimCtrl, 101113633). 
Views and opinions expressed are however those of the authors only and do not necessarily reflect those of the European Union or the European Research Council Executive Agency.

\section*{Data availability} 

The data and codes associated with this manuscript version are available under DOI: \href{https://doi.org/10.5281/zenodo.20442086}{10.5281/zenodo.20442086}.

\begin{appendix}

\section{Details of models and numerical simulations} 
\label{app:sec:numerics}

\subsection{Integrable Ising chain} 
\label{app:sec:TFIC}

\paragraph*{Fermionic representation.} The integrable transverse-field Ising Hamiltonian, Eq.~\eqref{eq:H_TFIC}, can be Jordan-Wigner-transformed to fermionic operators as~\cite{Suzuki2013Quantum}
\begin{multline}
    H= -\sum_j \left[ c_j^\dagger c_{j+1}^\dagger + c_j^\dagger c_{j+1}^\phdagger - c_j^\phdagger c_{j+1}^\dagger \right. \\ 
    \left. - c_j c_{j+1} - 2 g c_j^\dagger c_j^\phdagger \right] - Lg,
\end{multline}
where the periodic boundary conditions fix $c_{j+L} = -c_j$ (even fermionic number sector). By translation invariance, the Hamiltonian can be diagonalized by a Fourier transform. It is convenient to use the convention
\begin{equation}
    \label{app:eq:Fourier_convention}
	\tilde{c}_k := \frac{e^{i\pi/4}}{\sqrt{L}} \sum_j e^{-i k j} c_j,
\end{equation}
with momenta (for the chosen parity sector, and a chain of even length)
\begin{equation}
    k = \frac{(2n+1)\pi}{L}, \qquad \text{for } n =-\frac{L}{2}, -\frac{L}{2}+1,\dots, \frac{L}{2}-1.
\end{equation}
One finds
\begin{equation}
    \label{app:eq:H_TFIC_Fourier}
	H = 2 \sum_{k>0} \psi_k^\dagger \left[ (\lambda-\cos k)\tau^z + \sin k \, \tau^x \right] \psi_k,
\end{equation}
where it was conveniently grouped
\begin{equation}
	\psi_k^\dagger := 
	\begin{pmatrix}
		\tilde{c}_k^\dagger 	&\tilde{c}_{-k}^\phdagger
	\end{pmatrix}, \qquad
	\psi_k := 
	\begin{pmatrix}
		\tilde{c}_k \\
		\tilde{c}_{-k}^\dagger
	\end{pmatrix}, 
\end{equation}
and $\tau^{x,y,z}$ are a set of Pauli matrices that act on the internal indices of the vectors $\psi_k^\phdagger,\psi_k^\dagger$. From Eq.~\eqref{app:eq:H_TFIC_Fourier}, one can see that the transverse-field Ising Hamiltonian has been reduced to a collection of $L/2$ non-interacting two-level systems.

\paragraph*{Adiabatic gauge potential.} The AGP of the transverse-field Ising chain follows from that of the two-level systems in momentum space~\cite{delCampo2012Assisted}:
\begin{equation}
    A_\text{GS}(\lambda) = - \sum_{k> 0} \frac{\sin k}{2(1+\lambda^2-2\lambda \cos k)} \psi_k^\dagger \tau^y \psi_k.
\end{equation}
To transform back to real space, one first needs to expand 
\begin{equation}
    A_\text{GS}(\lambda) = i\sum_{k> 0} \frac{\sin k}{2(1+\lambda^2-2\lambda \cos k)} \left[ \tilde{c}_k^\dagger \tilde{c}_{-k}^\dagger - \tilde{c}_{-k} \tilde{c}_k \right],
\end{equation}
and then using Eq.~\eqref{app:eq:Fourier_convention}
\begin{align}
    A_\text{GS}(\lambda) &= \frac{1}{L} \sum_{k>0} \sum_{jl} \frac{\sin k \, e^{ikj-ikl}}{2(1+\lambda^2-2\lambda\cos k)}  \left[ c_j^\dagger c_l^\dagger + c_j c_l  \right] \\
    &= \frac{i}{L} \sum_{k>0} \sum_{j>l} \frac{\sin k \, \sin[k(j-l)]}{1+\lambda^2-2\lambda\cos k} \left[ c_j^\dagger c_l^\dagger + c_j c_l  \right],
\end{align}
having used the anticommutation of $c$'s and restricted the sum to $j>l$ ($j=l$ is irrelevant since the operator vanishes). We now need to evaluate the sum
\begin{align}
    I &\equiv \frac{i}{L} \sum_{k>0} \frac{\sin k \, \sin(kr) }{1+\lambda^2-2\lambda\cos k}\\
    &\approx i\int_0^\pi \frac{dk}{2\pi} \frac{\sin k \, \sin(kr) }{1+\lambda^2-2\lambda\cos k}\\
    &= \int_{-\pi}^\pi \frac{dk}{4\pi} \frac{\sin k \, e^{ikr} }{1+\lambda^2-2\lambda\cos k}
\end{align}
with $r>0$. We perform the substitution $z \equiv e^{ik}$, finding
\begin{equation}
    I = -\oint_{|z|=1} \frac{dz}{8\pi \lambda} \, \frac{(z^2-1)z^{r-1}}{(z-1/\lambda)(z-\lambda)}.
\end{equation}
Now one can use the residue theorem. Focusing on the region encircled by the integration contour, and if $\lambda>1$, there is a simple pole at $z=1/\lambda$: it follows
\begin{equation}
    I\big|_{\lambda>1} = \frac{1}{4i}\lambda^{-r-1}.
\end{equation}
Notice that there is no residue at $z=0$ since $r>0$. If instead $0<\lambda<1$, there is a simple pole at $z=\lambda$, and
\begin{equation}
    I\big|_{\lambda<1} = \frac{1}{4i}\lambda^{r-1} .
\end{equation}
The results for $\lambda<0$ are the same (i.e.\ one can just replace $\lambda \to |\lambda|$). One concludes that 
\begin{equation}
    A = \frac{i}{4} \sum_{l>j} \lambda^{(l-j)\mathrm{sgn}(1-|\lambda|) -1} \left[ c_j^\dagger c_l^\dagger + c_j c_l  \right].
\end{equation}

The last step is to Jordan-Wigner-transform back to spins: one finds exactly Eq.~\eqref{eq:TFIC_AGP}. Notice that we are disregarding boundary terms for simplicity, see Ref.~\cite{delCampo2012Assisted}.

\paragraph*{Norm of the gauge potential.} The norm squared of the AGP can be found from~\cite{damski2013fidelity,luo2018fidelity}
\begin{align}
    \| A_\text{GS}(\lambda) \|^2 &= \sum_{k>0} \frac{\sin^2 k}{2(1+\lambda^2-2\lambda \cos k)^2} \\
    &\approx \frac{L}{4\pi} \int_0^\pi \frac{\sin^2 k}{(1+\lambda^2-2\lambda \cos k)^2}\, dk \\
    &= \frac{iL}{32\pi \lambda^2} \oint \frac{(z^2-1)^2}{z(z-\lambda)^2(z-1/\lambda)^2} \, dz,
\end{align}
having performed again the substitution $z \equiv e^{ik}$. There are three poles now: a single pole at $z=0$, with residue 1, that needs to be always taken; a double pole at $z=\lambda$, with residue $(\lambda^2+1)/(\lambda^2-1)$, to be taken when $|\lambda|<1$; and a double pole at $z=1/\lambda$, with residue $(\lambda^2+1)/(1-\lambda^2)$, to be taken when $|\lambda|>1$. Putting everything together, one finds
\begin{equation}
    \label{app:eq:norm_AGP_TFIC}
    \| A_\text{GS}(\lambda) \|^2=
    \begin{cases}
        \frac{L}{8(1-\lambda^2)}     & \lambda<1 \\
        \frac{L}{8\lambda^2(\lambda^2-1)}  & \lambda>1,
    \end{cases}
\end{equation}
which is Eq.~\eqref{eq:norm_AGP_TFIC}.

\paragraph*{Constant-speed parametrization.} One can check that 
\begin{equation}
\label{app:eq:cs_sol_TFIC}
    \lambda_\mathrm{cs}(t) = 
    \begin{cases}
        \displaystyle
        \frac{\tan[v_<(t-t_<)]}{\sqrt{1+\tan^2[v_<(t-t_<)]}} &\lambda\leq 1 \\[12pt]
        \displaystyle
        \sqrt{1+\tan^2[v_>(t-t_>)]} &\lambda\geq 1,
    \end{cases}
\end{equation}
is the solution to the differential equation~\eqref{eq:constant-speed-condition}. To fix the integration constants $v_<,v_>,t_<,t_>$, we require that the schedule obey the endpoint conditions
\begin{equation}
    \lambda(0) = \lambda_1 > 1, \qquad
    \lambda(T) = \lambda_2 = 0.
\end{equation}
We also impose $v_< = v_>$, i.e.\ that the speed is indeed constant: Eq.~\eqref{app:eq:cs_sol_TFIC} reduces then to Eq.~\eqref{eq:cs_sol_TFIC}.

The conditions above fix
\begin{subequations}
    \label{app:eq:TFIC_integration_constants}
    \begin{gather}
        v_> = -\frac{1}{t_c} \arctan \sqrt{\lambda_1^2-1}, \quad t_> = t_c, \quad
        \text{for } t<t_c, \\
        \label{eq:v_RC}
        v_< = -\frac{\pi}{2(T-t_c)}, \quad t_< = T, \quad
        \text{for } t>t_c,
    \end{gather}
\end{subequations}
where $t_c$ is the time at which $\lambda$ crosses $\lambda_c=1$, and has value
\begin{equation}
    \label{app:eq:TFIC_t_c}
    t_c = \frac{2T\arctan \sqrt{\lambda_1^2-1}}{\pi + 2\arctan \sqrt{\lambda_1^2-1}}.
\end{equation}

\subsection{Cluster Ising chain} 
\label{app:sec:CIM}

\paragraph*{Fermionic representation.} Using the same momentum-space basis of the transverse-field Ising chain, the cluster Ising Hamiltonian reads
\begin{multline}
	H = 2 \sum_{k>0} \psi_k^\dagger \big[ (\mu_2 \cos k+ \mu_3\cos 2k - \mu_1)\tau^z \\
    - (\mu_2\sin k+ \mu_3\sin 2k) \, \tau^x \big] \psi_k.
\end{multline}
One can check that the AGP for the path in Eq.~\eqref{eq:CIM_path} reads
\begin{equation}
    A_\text{GS}(\lambda) = - \sum_{k> 0} \frac{\sin k}{1+\lambda^2+(\lambda^2-1) \cos k} \psi_k^\dagger \tau^y \psi_k.
\end{equation}

\paragraph*{Adiabatic gauge potential.} Proceeding as done for the transverse-field Ising chain, one determines
\begin{equation}
    A_\text{GS}(\lambda) = \frac{2i}{L} \sum_{k>0} \sum_{j>l} \frac{\sin k \, \sin[k(j-l)]}{1+\lambda^2+(\lambda^2-1)\cos k} \left[ c_j^\dagger c_l^\dagger + c_j c_l  \right]
\end{equation}
and thus the integral to evaluate is 
\begin{equation}
    I = \int_{-\pi}^\pi \frac{dk}{2\pi} \frac{\sin k \, e^{ikr} }{1+\lambda^2 + (\lambda^2-1) \cos k}.
\end{equation}
The substitution $z \equiv e^{ik}$ brings it to the form
\begin{equation}
    I = -\oint_{|z|=1} \frac{dz}{2\pi \lambda} \, \frac{(z^2-1)z^{r-1}}{2z(\lambda^2+1) + (\lambda^2-1)(z^2+1)},
\end{equation}
which can be evaluated via the residue theorem. The integration contour encircles the poles
\begin{equation}
    z= -\frac{\lambda-1}{\lambda+1}, \qquad \text{ for } \lambda>0
\end{equation}
and
\begin{equation}
    z= -\frac{\lambda+1}{\lambda-1}, \qquad \text{ for } \lambda<0.
\end{equation}
This leads to
\begin{equation}
    I\big|_{\lambda>0} = \frac{-i}{\lambda^2-1} \left(\frac{1-\lambda}{1+\lambda} \right)^r, \quad 
    I\big|_{\lambda<0} = \frac{-i}{\lambda^2-1} \left(\frac{1+\lambda}{1-\lambda} \right)^r.
\end{equation}
Finally,
\begin{equation}
    A = i \sum_{l>j} f_{l-j}(\lambda) \left[ c_j^\dagger c_l^\dagger + c_j c_l  \right]
\end{equation}
with
\begin{equation}
    f_r(\lambda) = \frac{1}{\lambda^2-1}
    \begin{cases}
        \left(\frac{1-\lambda}{1+\lambda} \right)^r & \lambda>0 \\
        \left(\frac{1+\lambda}{1-\lambda} \right)^r & \lambda<0 ,
    \end{cases}
\end{equation}
and in terms of Pauli matrices:
\begin{multline}
    A_\text{GS}(\lambda) =
    \frac{1}{2}\sum_{r \geq 1} f_{l-j}(\lambda) \sum_j \left( \sigma_j^z \sigma_{j+1}^x \cdots \sigma_{j+r-1}^x \sigma_{j+r}^y \right. \\
    \left. + \sigma_j^y \sigma_{j+1}^x \cdots \sigma_{j+r-1}^x \sigma_{j+r}^z \right).
\end{multline}

\paragraph*{Norm of the gauge potential.} One can find the AGP norm, in the thermodynamic limit, from
\begin{align}
    \| A_\text{GS}(\lambda) \|^2 &= 2\sum_{k> 0} \frac{\sin^2 k}{[1+\lambda^2+(\lambda^2-1) \cos k]^2}  \\
    &\approx \frac{L}{\pi} \int_0^\pi \frac{\sin^2 k}{[1+\lambda^2+(\lambda^2-1) \cos k]^2} dk \\
    &= \frac{iL}{2\pi} \oint \frac{(z^2-1)^2}{z[2z+2z\lambda^2 + (\lambda^2-1)(z^2+1)]^2} \, dz,
\end{align}
having used again $z\equiv e^{ik}$. The residue theorem leads to
\begin{equation}
    \| A_\text{GS}(\lambda) \|^2 = \frac{L}{2 |\lambda| (1+|\lambda|)^2}.
\end{equation}

\subsection{Non-integrable transverse-field Ising chain} 
\label{app:sec:NITFIC}

As stated in the main text, the treatment of the integrable TFIC hinged upon the knowledge of the exact expressions for the ground-state AGP. Here, we show instead that it is possible to obtain the constant-speed parametrization also fully numerically. In order to reach large system sizes, we decided to use tensor networks; any other numerical technique that grants access to both time evolution and the instantaneous ground state would work as well. 

The first step consists in determining the shape of the curve $\lambda_\mathrm{cs}(t)$. By Eq.~\eqref{eq:constant-speed-implicit}, $\lambda_\mathrm{cs}(t)$ follows from the knowledge of $\norm{A_\text{GS}(\lambda,L)}$, which in turn can be computed from the ground state wavefunction. Therefore, we first obtain the exact ground state of the Hamiltonian via the density-matrix renormalization group (DMRG)~\cite{White1992Density,*White1993Density}, for each value of $\lambda$ on a discrete grid. We employ a precision high enough, that is not saturated even at the finite-size precursor of the critical point ($\chi_\mathrm{DMRG} \leq 200$).
Then, we obtain $A_\text{GS}(\lambda)$ via Eq.~\eqref{eq:AGP_GS} using the outer product of matrix product states and finite derivatives.
Computing its norm, we can obtain the constant-speed curve $\lambda_\mathrm{cs}(t)$ by solving numerically Eq.~\eqref{eq:constant-speed-implicit}.
We finally compress $A_\text{GS}(\lambda)$ to a lower operator bond dimension $\chi_\mathrm{AGP} \ll \chi_\mathrm{DMRG}^2$ (in practice, $\chi_\mathrm{AGP} \leq 50$), to keep the algorithm fast enough. 
We checked that such a strong compression does not alter significantly the final results. Finally, we simulate the evolution protocols via the time-dependent variational principle (TDVP)~\cite{Haegeman2011Time}. For simulation efficiency (especially for the CD-assisted schedules, which require a large-bond-dimension Hamiltonian), we impose a rather strong cutoff on the bond dimension: $\chi_\mathrm{TDVP} \leq 100$. 

As a last comment, notice that, in the integrable case, the curve $\lambda_\mathrm{cs}(t)$ was obtained for a thermodynamically large system and then used for time-evolving also finite systems. This is clearly not doable in the non-integrable case. We thus resort to computing $\lambda_\mathrm{cs}(t)$ for a system size ($L=400$) far exceeding the system sizes considered for evolution ($L\leq100$), and with higher precision ($\chi_\mathrm{DMRG} \leq 400$).
The curve $\lambda_\mathrm{cs}(t)$ obtained with the larger system acts effectively as the thermodynamic limit for the smaller systems that were time-evolved.

\subsection{Anisotropic XY Lipkin-Meshkov-Glick model}

We consider the LMG Hamiltonian in Eq.~\eqref{eq:LMG_H-1st_order}. Following \cite{dusuel2005continuous}, in the $L\gg 1$ limit, we can study the first order quantum corrections by means of an Holstein-Primakoff representation of spin operators. First of all we perform a rotation of the spin operators, that brings the $z$ axis along the semiclassical magnetization.
\begin{align}
    \begin{pmatrix}S_x\\
    S_y\\
    S_z
    \end{pmatrix} = \begin{pmatrix}\cos\theta\cos\phi &-\sin\phi &\sin\theta\cos\phi\\
    \cos\theta\sin\phi &\cos\phi &\sin\theta\sin\phi\\
    -\sin\theta &0 &\cos\theta\end{pmatrix}\begin{pmatrix}\tilde{S}_x\\
    \tilde{S}_y\\
    \tilde{S}_z
    \end{pmatrix},
\end{align}
where the rotation angles depends on the $\lambda$ and $g$ parameters as follows:
\begin{align}
    \theta &= \begin{cases}
        0 &g\geq\frac{1+|\lambda|}{2}\\
       \arccos\left[\frac{2g}{1+\abs{\lambda}}\right] &0\leq g<\frac{1+|\lambda|}{2}
    \end{cases}\,,\\
    \phi &= \begin{cases}
        0~\mathrm{or}~\pi &\lambda\geq 0\\
        \pi/2~\mathrm{or}~3\pi/2 &\lambda< 0
    \end{cases}\,,
\end{align}
where the two possibilities for the choice of the $\phi$ angle reflect the ground-state degeneracy. 

The Holstein-Primakoff representation is then applied to the rotated spin operators
\begin{subequations}
\begin{align}
    \tilde{S}_z &= L/2-a^\dagger a,\\
    \tilde{S}_+ &= \sqrt{L} \big(1-a^\dagger a/L \big)^{1/2}a, \\
    \tilde{S}_- &= \sqrt{L}\ a^\dagger \big(1-a^\dagger a/L\big)^{1/2},
\end{align}
\end{subequations}
where $\tilde{S}_{\pm}= \tilde{S}_x\pm i\tilde{S}_y$.
Inserting these expressions in the Hamiltonian written in terms of the rotated spin operators, and expanding the argument of the square roots and keeping terms of order $L$, $\sqrt{L}$, and $L^0$ yields
\begin{align}
    H &\approx L\varepsilon_0+(1+\abs{\lambda})(1-m^2)/4\notag\\
    &\phantom{=} +\left[gm+3(1+\abs{\lambda})(1-m^2)/2-1\right] a^\dagger a\notag\\
    &\phantom{=} +\left[(1-\abs{\lambda})/4-(1+\abs{\lambda})m^2/4\right]\big(a^{\dagger 2}+a^2 \big),
\end{align}
where $\varepsilon_0 = -(1+\abs{\lambda})(1-m^2)/4-gm$ is the mean-field ground-state energy per spin and $m = \cos\theta$. 

This quadratic Hamiltonian can be diagonalized by a Bogoliubov transformation
\begin{subequations}
\begin{align}
    a^\dagger &= \cosh(\Theta/2)b^\dagger + \sinh(\Theta/2)b,\\
    a &= \sinh(\Theta/2)b^\dagger+\cosh(\Theta/2)b,
\end{align}    
\end{subequations}
with
\begin{align}
     \tanh\Theta = \begin{cases}
        \displaystyle \frac{4g^2+\lambda^2-1}{3\lambda^2+4\abs{\lambda}+1-4g^2} &g\geq \frac{1+\abs{\lambda}}{2}\\
        \displaystyle \frac{\abs{\lambda}}{2g-1} &0\leq g<\frac{1+\abs{\lambda}}{2}.
    \end{cases}
\end{align}
This leads to the diagonal form of the Hamiltonian
\begin{align}
    H\approx L\varepsilon_0+\delta e + \omega b^\dagger b,
\end{align}
where
\begin{align}
    \delta e = \begin{cases}
       -g+\frac{1}{2}+\frac{1}{2}\sqrt{(2g-1)^2-\lambda^2}\ &g\geq \frac{1+\abs{\lambda}}{2}\\\\
    -\frac{\abs{\lambda}}{2}+\frac{1}{2}\sqrt{\frac{2\abs{\lambda}}{1+\abs{\lambda}}\left[(1+\abs{\lambda})^2-4g^2\right]} &0\leq g<\frac{1+\abs{\lambda}}{2},
    \end{cases}
\end{align}
and
\begin{align}
    \omega = \begin{cases}
        \sqrt{(2g-1)^2-\lambda^2} &g\geq \frac{1+\abs{\lambda}}{2}\\\\
        \sqrt{\frac{2\abs{\lambda}}{1+\abs{\lambda}}\left[(1+\abs{\lambda})^2-4g^2\right]} &0\leq g<\frac{1+\abs{\lambda}}{2}.
    \end{cases}
\end{align}
We have thus mapped the time-independent LMG Hamiltonian to a single harmonic oscillator, up to subleading contributions in the $1/L$ expansion.

As shown in Eq.~\eqref{eq:AGP_norm_FS}, the squared norm of the ground-state AGP is given by the fidelity susceptibility $\chi_\mathrm{F}$, which in this case can be computed as~\cite{kwok2008quantum}
\begin{align}\label{eq:FS XY LMG}
    \chi_\mathrm{F}(\lambda) = \sum_{n\neq 0}\frac{\abs{\bra{n}(S_y^2-S_x^2)/L\ket{0}}^2}{(E_n-E_0)^2}.
\end{align}
The low-energy spectrum of the model is mapped, in the $L\gg 1$ limit, to the spectrum of a simple harmonic oscillator. The eigenstates are $\{\ket{n}\}$, where $b^\dagger b\ket{n} = n\ket{n}$ and the eivenvalues are $E_n-E_0 = \omega n$. 

Then, writing the driving Hamiltonian in terms of the rotated collective spin operators, and using the mapping to bosonic raising and lowering operators $b,b^\dagger$, one obtains
\begin{multline}
\frac{1}{L}(S_y^2-S_x^2) = -\frac{L}{4}\sin^2\theta+\frac{\sqrt{L}}{2}\sin{\theta}\cos{\theta}(b^\dagger+b)\\
+\frac{1}{4}\left[\sqrt{\frac{1-\tanh{\Theta}}{1+\tanh{\Theta}}}(b^\dagger-b)^2-\cos^2\theta\sqrt{\frac{1+\tanh{\Theta}}{1-\tanh{\Theta}}}(b^\dagger+b)^2\right]
\end{multline}
Inserting the expression above into the fidelity susceptibility, Eq.~\eqref{eq:FS XY LMG}, the only non vanishing terms when evaluating $S_y^2-S_x^2$ between the states $\bra{n}$ and $\ket{0}$ are the ones proportional to $b^\dagger$ and $b^{\dagger 2}$, leading to
\begin{multline}
\chi_\mathrm{F}(\lambda) = \frac{L}{4\omega^2}\cos^2\theta\sin^2\theta\\
+ \frac{1}{32\omega^2}\left[\sqrt{\frac{1-\tanh{\Theta}}{1+\tanh{\Theta}}}-\cos^2\theta\sqrt{\frac{1+\tanh{\Theta}}{1-\tanh{\Theta}}}\right]
\end{multline}
Specializing the expressions for $\theta$, $\tanh\Theta$ and $\omega$ in the two ferromagneti phases ($g<(1+\abs{\lambda})/2$), one obtains the expressions for the fidelity susceptibility in Eq.~\eqref{eq:FS_LMG-1st_order}.

\subsection{Ising Lipkin-Meshkov-Glick model}
\label{app:sec:LMG}

We consider the Ising Lipkin-Meshkov-Glick Hamiltonian in Eq.~\eqref{eq:LMG Hamiltonian}. Following \cite{dusuel2005continuous}, in the $L\gg 1$ limit, we can study the first order quantum corrections by means of an Holstein-Primakoff representation of spin operators. First of all we perform a rotation of the spin operators around the $y$ axis, that brings the $z$ axis along the semiclassical magnetization:
\begin{align}
    \begin{pmatrix}S_x\\
    S_y\\
    S_z
    \end{pmatrix} = \begin{pmatrix}\cos\theta &0 &\sin\theta\\
    0 &1 &0\\
    -\sin\theta &0 &\cos\theta
    \end{pmatrix}\begin{pmatrix}\tilde{S}_x\\
    \tilde{S}_y\\
    \tilde{S}_z
    \end{pmatrix},
\end{align}
where the rotation angle is given by 
\begin{align}
    \theta = \begin{cases}
        0 &\lambda\geq 1\\
        \arccos(\lambda) &0\leq \lambda<1 .
    \end{cases}
\end{align}
The Holstein-Primakoff representation is then applied to the rotated spin operators
\begin{subequations}
\begin{align}
    \tilde{S}_z &= L/2-a^\dagger a,\\
    \tilde{S}_+ &= \sqrt{L} \big(1-a^\dagger a/L \big)^{1/2}a, \\
    \tilde{S}_- &= \sqrt{L}\ a^\dagger \big(1-a^\dagger a/L\big)^{1/2},
\end{align}
\end{subequations}
where $\tilde{S}_{\pm}= \tilde{S}_x\pm i\tilde{S}_y$.
Inserting these expressions in the Hamiltonian written in terms of the rotated spin operators, and expanding the argument of the square roots and keeping terms of order $L$, $\sqrt{L}$, and $L^0$ yields
\begin{align}
    H\approx L\varepsilon_0 +
    \frac{1-m^2}{2}+ \big(2+2\lambda m-3m^2\big)a^\dagger a
    +\frac{m^2}{2} \big(a^{\dagger 2}+a^2 \big),
\end{align}
where $\varepsilon_0 = (-1-2\lambda m+m^2)/2$ is the mean-field ground-state energy per spin and $m = \cos\theta$. This quadratic Hamiltonian can be diagonalized by a Bogoliubov transformation
\begin{subequations}
\begin{align}
    a^\dagger &= \cosh(\Theta/2)b^\dagger + \sinh(\Theta/2)b,\\
    a &= \sinh(\Theta/2)b^\dagger+\cosh(\Theta/2)b,
\end{align}    
\end{subequations}
with
\begin{align}
    \tanh\Theta = \begin{cases}
        \displaystyle \frac{1}{2\lambda-1} &\lambda\geq 1\\\\
        \displaystyle \frac{\lambda^2}{2-\lambda^2} &0\leq \lambda<1.
    \end{cases}
\end{align}
This leads to the diagonal form of the Hamiltonian
\begin{align}
    H\approx L\varepsilon_0+\delta e + \omega b^\dagger b,
\end{align}
where
\begin{align}
    \delta e = \begin{cases}
        -\lambda+\frac{1}{2}+\sqrt{(\lambda-1)\lambda} &h\geq 1\\\\
        -\frac{1}{2}+\sqrt{1-\lambda^2} &0\leq \lambda<1,
    \end{cases}
\end{align}
and
\begin{align}
    \omega = \begin{cases}
        2\sqrt{(\lambda-1)\lambda} &h\geq 1\\\\
        2\sqrt{1-\lambda^2} &0\leq \lambda<1.
    \end{cases}
\end{align}
We have thus mapped the time-independent LMG Hamiltonian to a single harmonic oscillator, up to subleading contributions in the $1/L$ expansion.

Considering corrections up to $\mathcal{O}(L^{-2})$ the first two moments total spin operators \cite{dusuel2005continuous} in the symmetric phase ($\lambda>1$) read
\begin{subequations}
\label{app:eq:lmg_exp_fm}
\begin{align}
    \frac{2\langle S_z\rangle}{L} &= 1+\frac{1}{L}\left[1-\frac{1/2 - \lambda}{2\sqrt{\lambda(\lambda-1)}}\right]+\mathcal{O}(L^{-2})\\
    \frac{2\langle S_x\rangle}{L} &= \frac{2\langle S_y\rangle}{L} = 0\\
    \frac{4\langle S_z^2\rangle}{L^2} &= 1+\frac{1}{L}\left[2+\frac{1-2\lambda}{\sqrt{\lambda(\lambda-1)}}\right]+\mathcal{O}(L^{-2})\\
    \frac{4\langle S_x^2\rangle}{L^2} &= \frac{1}{L}\sqrt{\frac{\lambda}{\lambda-1}}+\mathcal{O}(L^{-2})\\
    \frac{4\langle S_y^2\rangle}{L^2} &= \frac{1}{L}\sqrt{\frac{\lambda-1}{\lambda}}+\mathcal{O}(L^{-2})
\end{align}
\end{subequations}
while in the broken phase ($0\leq\lambda\leq 1$) we have
\begin{subequations}
\label{app:eq:lmg_exp_pm}
\begin{align}
    \frac{2\langle S_x\rangle}{L} &= \sqrt{1-\lambda^2}+\frac{1}{L}\left[\frac{1}{\sqrt{1-\lambda^2}}+\frac{2+\lambda^2}{2(1-\lambda^2)}\right]+\mathcal{O}(L^{-2})\\
    \frac{2\langle S_y\rangle}{L} &= \mathcal{O}(L^{-2})\\
    \frac{2\langle S_z\rangle}{L} &= \lambda+\frac{1}{L}\frac{\lambda}{\sqrt{1-\lambda^2}}\\
    \frac{4\langle S_x^2\rangle}{L^2} &= 1-\lambda^2+\frac{1}{L}\left[2-\frac{2}{\sqrt{1-\lambda^2}}\right]+\mathcal{O}(L^{-2})\\
   \frac{4\langle S_y^2\rangle}{L^2} &= \frac{1}{L}\sqrt{1-\lambda^2}+\mathcal{O}(L^{-2})\\
    \frac{4\langle S_z^2\rangle}{L^2}&= \lambda^2+\frac{1}{L}\left[\frac{1+\lambda^2}{\sqrt{1-\lambda^2}}\right]+\mathcal{O}(L^{-2})
\end{align}
\end{subequations}
These expressions uniquely characterize the Gaussian ground state of the LMG model. Crucially, these expressions are discontinuous at the phase transition point ($\lambda=1$), thus implying the discontinuity of the ground state manifold in agreement with the infinite geometric length.

As shown in Eq.~\eqref{eq:AGP_norm_FS}, the squared norm of the ground-state AGP is given by the fidelity susceptibility $\chi_\mathrm{F}$, which in this case can be computed as~\cite{kwok2008quantum}
\begin{align}\label{eq:FS LMG}
    \chi_\mathrm{F}(\lambda) = \sum_{n\neq 0}\frac{\abs{\bra{n}2S_z\ket{0}}^2}{(E_n-E_0)^2}.
\end{align}
The low-energy spectrum of the model is mapped, in the $L\gg 1$ limit, to the spectrum of a simple harmonic oscillator. The eigenstates are $\{\ket{n}\}$, where $b^\dagger b\ket{n} = n\ket{n}$ and the eivenvalues are $E_n-E_0 = \omega n$. 

Then, writing the driving Hamiltonian in terms of the rotated collective spin operators, and using the mapping to bosonic raising and lowering operators $b,b^\dagger$, one obtains
\begin{multline}
    2S_z = \sin\theta\sqrt{L}\left(\frac{1+\tanh\Theta}{1-\tanh\Theta}\right)^{1/4} \big( b^\dagger+b \big) -\cos\theta(L+1)\\
    +\cos\theta\left[\frac{1+2b^\dagger b}{\sqrt{1-\tanh^2\Theta}}+\frac{\tanh\Theta}{2\sqrt{1-\tanh^2\Theta}} \big(b^{\dagger 2}+b^2 \big)\right].
\end{multline}
Inserting the expression above into the fidelity susceptibility, Eq.~\eqref{eq:FS LMG}, the only non vanishing terms when evaluating $S_z$ between the states $\bra{n}$ and $\ket{0}$ are the ones proportional to $b^\dagger$ and $b^{\dagger 2}$, leading to
\begin{align}
\chi_\mathrm{F}(\lambda) = \frac{L\sin^2\theta}{\omega^2} \sqrt{\frac{1+\tanh\Theta}{1-\tanh\Theta}}+\frac{\cos^2\theta\tanh^2\Theta}{2\omega^2(1-\tanh^2\Theta)}.
\end{align}
Specializing the expressions for $\theta$, $\tanh\Theta$ and $\omega$ in the two phases, one obtains the expressions for the fidelity susceptibility in Eqs.~\eqref{eq:FS LMG paramagnetic} and~\eqref{eq:FS LMG symmetric}.

\section{Algorithmic procedure for finite-speed schedule of unknown traversable QPT point}
\label{app:sec:algorithm}

We now describe an operational definition of traversability and the finite-speed schedule that does not require prior knowledge of the transition point's location. 
The crucial insight is that the schedule is entirely local in $\lambda$: at each step, only the current ground state is needed to determine the next, so the critical point is discovered rather than assumed.
This is possible because the adiabatic gauge potential $A_\text{GS}(\lambda)$ is well-defined at any fixed $\lambda$, and the constant-speed schedule is generated by a simple ODE that can be stepped forward without global knowledge of the phase diagram.

The algorithm proceeds as follows---assuming finite system size $L$, fixed small step size $\Delta t$ and a speed $v$.
At time $t$, the system is at $\lambda = \lambda_\text{cs}(t)$ with ground state $\ket{\psi_\text{GS}[\lambda]}$.
From this state, one computes the adiabatic gauge potential $A_\text{GS}(\lambda)$, for instance via the variational principle; crucially, this can be done locally in $\lambda$.
The key observation to obtain the finite-speed schedule is its equivalence to the intrinsic length parametrization of the parameter manifold (see Eq.~\eqref{eq:intrinsic-length}), which can be evaluated locally from the ground state alone.
Consequently, the parameter is updated according to
\begin{equation}
    \lambda_\text{cs}(t + \Delta t) - \lambda_\text{cs}(t) = \frac{v \,\Delta t}{\|A_\text{GS}(\lambda)\|} \equiv \mathrm{d}\lambda,
\end{equation}
and the ground state is advanced as
\begin{equation}
    \ket{\psi_\text{GS}[\lambda + \mathrm{d}\lambda]} \approx \ket{\psi_\text{GS}[\lambda]} + \mathrm{d}\lambda\, A_\text{GS}(\lambda) \ket{\psi_\text{GS}[\lambda]}.
\end{equation}
This procedure is iterated until a target value $\lambda_1$ is reached.

The total time $T_\text{cs}(v)$ required to reach $\lambda_1$ is not fixed in advance and depends on the available resource $v$ as well as the geometry of the path.
At finite system size, the ground state energy is analytic in $\lambda$, so $\|A_\text{GS}(\lambda)\|$ remains bounded and the schedule always terminates in finite time for any $v > 0$.

However, the required time $T_\text{cs}(L)$ as a function of system size distinguishes traversable from nontraversable transitions: for a traversable QPT, $T_\text{cs}(L)$ converges to a finite value in the thermodynamic limit, as does the schedule acquired via the algorithm above.
In contrast, for a nontraversable QPT it diverges, signaling that no finite resource budget suffices to cross the transition.

The algorithm outlined above corresponds to a simple first order Runge-Kutta solution to the constant-speed ODE; to improve the accuracy of the algorithm higher order Runge-Kutta methods may be considered.
Additionally, when a tensor network or variational ansatz is considered, the updated state serves as an initial guess for the variational solver at $\lambda + \mathrm{d}\lambda$, rather than being used directly.

\section{Properties of adiabatic gauge potentials}
\label{app:Kato_vs_Hastings}

In this section, we discuss properties of the Kato (K) and Osborne-Hastings (OH) adiabatic gauge potentials (AGP) and compare the two. 
In a nutshell, we show that:
(i) the OH GS-AGP can be thought of as using the residual gauge freedom from the action on excited states, to trade Hilbert-Schmidt norm for locality;
(ii) the OH GS-AGP converges to the full Kato AGP (as opposed to the GS Kato AGP) as one approaches the phase transition point;
(iii) unlike the OH GS-AGP, which is quasilocal by construction, there are no guarantees for the locality properties of the GS Kato AGP; and
(iv) starting from the GS deep in the bulk of a phase, the traversability properties of CD evolution with the OH GS-AGP are the same as the traversability properties using the Kato GS-AGP. 

\subsection{Full adiabatic gauge potential}

Adiabatic gauge potentials $A$ are not gauge-invariant objects. They are solutions to the equation~\cite{Kolodrubetz2017Geometry}
\begin{equation}
\label{eq:AGP_defining_eq}
    \mathcal{L}^2(A)  + \mathcal{L}(\partial_\lambda H)=0,
\end{equation}
where $\mathcal{L}(\cdot)=i[H(\lambda),\cdot]$ is the Liouvillian superoperator. 
Different gauge choices correspond to adding or subtracting, from a fixed solution, terms that commute with the Hamiltonian; this makes the kernel of $\mathcal{L}$ nontrivial and $\mathcal{L}$ noninvertible. 

Out of all gauge choices, 
\begin{equation}
    A^\text{K}(\lambda) = \mathcal{L}^+(-\partial_\lambda H(\lambda)),
\end{equation}
corresponds to the minimum-norm solution to the defining equation~\eqref{eq:AGP_defining_eq}, where $\mathcal{L}^+$ denotes the Moore-Penrose pseudo-inverse of $\mathcal{L}$. 
The Kato (or parallel transport) gauge coincides with the minimum-norm choice among all possible gauge choices. This follows directly from the properties of the Moore-Penrose pseudo-inverse. 

The full Kato AGP achieves transitionless driving w.r.t.~\textit{all} eigenstates $\ket{n[\lambda]}$ of the control hamiltonian $H(\lambda)$. 
The following present three equivalent expressions for it: 
\begin{subequations}
\label{eq:full_Kato_AGP}
\begin{align}
    A^\text{K}(\lambda) &= \frac{1}{2i} \sum_{n} [\partial_\lambda \Pi_n[\lambda], \Pi_n[\lambda]] \label{eq:full_Kato_AGP-proj} \\
    &= -i \sum_{m\neq n\geq 0}\frac{\ket{n[\lambda]}\bra{n[\lambda]}\partial_\lambda H(\lambda)\ket{m[\lambda]}\bra{m[\lambda]} } {E_n(\lambda)-E_m(\lambda)} \label{eq:full_Kato_AGP-spec} \\
    &= i\lim_{\delta\to 0^+} \int_{-\infty}^\infty \mathrm d s\; \mathrm e^{-\delta |t|} F(s)\; \mathrm e^{i s H(\lambda)}\partial_\lambda H(\lambda) \mathrm e^{-isH(\lambda)}, \label{eq:full_Kato_AGP-int}
\end{align}
\end{subequations}
which we now briefly discuss:

(i) In Eq.~\eqref{eq:full_Kato_AGP-proj}, $\Pi_n[\lambda]=\dyad{n[\lambda]}{n[\lambda]}$ denotes the eigenstate projector. This definition generalizes in a straightforward way to degenerate subspaces by redefining the corresponding projectors. 

(ii) Equation~\eqref{eq:full_Kato_AGP-spec} corresponds to the expansion in the eigenbasis of the Hamiltonian, and corresponds to the standard AGP expression by Berry~\cite{Berry2009Transitionless}. We will use this definition below to argue about the norm density of $A^\text{K}$. 

(iii) Equation~\eqref{eq:full_Kato_AGP-int} is an integral representation of the Kato AGP using the Heisenberg picture, where $F(s)$ is a filter function\footnote{Adding a suitable regularization is required to make the $s$-integral convergent.}. 
We note that $s$ here is an integration parameter, and does not play the role of physical time. It is straightforward to see that the parallel-transport gauge corresponds to the condition that the Fourier transform $\tilde F(\omega)$ vanishes at $\omega=0$ (i.e.,\ $F(s)=-F(-s)$ is odd); 
indeed, inserting two resolutions of the identity in the eigenbasis of the Hamiltonian, one recovers the expression from the middle line using the choice $\tilde F(\omega) = -1/\omega$ where $\tilde F(\omega)$ is the Fourier transform of $F(s)$.

As discussed in the main text, for generic nonintegrable systems, the energy difference denominator in Eq.~\eqref{eq:full_Kato_AGP-spec} vanishes exponentially in the system size for eigenstates in the middle of the spectrum (so-called infinite-temperature states for spin and fermion systems). As a result, the full Kato AGP is a nonlocal operator~\cite{Saberi2014Adiabatic,Kolodrubetz2017Geometry,Sels2017Minimizing} with a divergent Hilbert-Schmidt norm density in the thermodynamic limit, i.e.,\ $L^{-d/2}\|A^\text{K}\|\to\infty$ as $L\to\infty$. This nonlocality is present already deep within the bulk of any quantum phase of matter, and is unrelated to QPTs. 

\subsection{Ground-state adiabatic gauge potentials}

In the main text, we discussed that, for gapped local Hamiltonians, one can define counterdiabatic driving in the thermodynamic limit, using the ground-state AGP:
\begin{eqnarray}
\label{eq:GS_Kato_AGP}
    A_\text{GS}^\text{K}(\lambda) &=& \frac{1}{i} [\partial_\lambda \Pi_\text{GS}[\lambda] , \Pi_\text{GS}[\lambda]]\\
    &=& -i \sum_{n>0}\frac{\ket{n[\lambda]}\bra{n[\lambda]}\partial_\lambda H(\lambda)\ket{0[\lambda]}\bra{0[\lambda]} } {E_n(\lambda)-E_0(\lambda)} + \mathrm{h.c.}, \nonumber
\end{eqnarray}
with $\Pi_\text{GS}[\lambda]$ the projector onto the ground-state $\ket{\psi_\text{GS}}\equiv\ket{0}$. 

For gapped systems, the GS Kato AGP has a finite Hilbert-Schmidt norm density. To see this, note first that the norm squared of the GS Kato AGP $\|A_\text{GS}\|^2 {=} \chi_F$ is equal to the fidelity susceptibility. For a perturbation given by the sum of local operators, $\partial_\lambda H = \sum_r h_r$, we have
\begin{eqnarray}
    \chi_F &=& \sum_{n>0}\frac{|\bra{n}\partial_\lambda H\ket{0}|^2 } {(E_n-E_0)^2}
    \leq \Delta^{-2} \bra{0} (\partial_\lambda H)^2 \ket{0}_c \nonumber\\
    &=& \Delta^{-2}\sum_{r,r'}\bra{0} h_r h_{r'}\ket{0}_c
    = C \Delta^{-2}\sum_{r,r'} \mathrm e^{-|r-r'|/\xi} \nonumber\\
    &=& C'(\xi, \Delta) L^{d},
\end{eqnarray}
where we used that the connected correlation function of a local operator decays exponentially in gapped phases, $\bra{0} h_r h_{r'}\ket{0}_c {\sim} e^{-|r-r'|/\xi}$, at a lengthscale set by the correlation length $\xi$. Importantly, since the constant $C'$ is independent of the linear system size $L$, we find that $L^{-d/2}\|A_\text{GS}\|^2 {\to} \text{const.}$ in the thermodynamic limit, and hence $A_\text{GS}$ has a finite Hilbert-Schmidt norm density.

When we focus on the ground state only, additional freedom arises from the action of $A_\text{GS}$ within the excited-state manifold, which remains unspecified. In general, we can write 
\begin{eqnarray}
    A_\text{GS} &=& 
    \begin{pmatrix}
        \Pi_\text{GS} A_\text{GS} \Pi_\text{GS}     & \Pi_\text{GS} A_\text{GS} (1-\Pi_\text{GS}) \\
        (1-\Pi_\text{GS}) A_\text{GS} \Pi_\text{GS} & (1-\Pi_\text{GS}) A_\text{GS} (1-\Pi_\text{GS})   
    \end{pmatrix}
    \nonumber\\ 
    &=& 
    \begin{pmatrix}
         A_\text{GS}^\text{gg}     &  A_\text{GS}^\text{eg} \\
        A_\text{GS}^\text{ge} & A_\text{GS}^\text{ee}
    \end{pmatrix}
\end{eqnarray}
where the $A_\text{GS}^\text{gg}$ diagonal entry corresponds to the ground-state manifold, while the $A_\text{GS}^\text{ee}$ entry to the excited state manifold; transitions in and out of the ground state manifold are given by the off-diagonal entries $A_\text{GS}^\text{eg}= (A_\text{GS}^\text{ge} )^\dagger$. 
The parallel-transport gauge for the ground state fixes $A_\text{GS}^\text{gg}=0$, whereas transitionless driving of the ground state fixes uniquely the off-diagonal elements $A_\text{GS}^\text{eg}$ connecting the GS to the excited states. The remaining matrix $A_\text{GS}^\text{ee}$ then describes precisely the extra freedom left; physically, this reflects the irrelevance of the transitions among excited states for the adiabatic dynamics of the ground state. 

Osborne and Hastings proposed to use this extra freedom to define a local GS AGP~\cite{osborne2007simulating,hastings2010locality}. To do this, they used the integral representation from Eq.~\eqref{eq:full_Kato_AGP-int} and specified the filter function  
\begin{eqnarray}
\label{eq:exact_filter}
    \tilde F_\text{OH}(\omega) = -\frac{1-\tilde g(\omega)}{\omega},
\end{eqnarray}
where $\tilde g(\omega)$ is a compactly supported function on the interval $[-\Delta,\Delta]$ set by the ground state gap $\Delta$, such that $\tilde g(0)=1$ and $\tilde g|_{|\omega|\geq\Delta}\equiv 0$\footnote{E.g.,\ the mollifier $\tilde g(\omega){=}\exp(1{-}(1{-}(\omega/\Delta)^{2})^{-1})$.}: 
\begin{eqnarray}
\label{eq:GS_HO_AGP}
    A^\text{OH}_\text{GS} &=&  i\int_{-\infty}^\infty \mathrm d s\;  F_\text{OH}(s)\; \mathrm e^{i s H}\partial_\lambda H \mathrm e^{-isH} \\
    &=& -i \sum_{m\neq n\geq 0} \tilde F_\text{OH}(E_n-E_m) \ket{n}\bra{n}\partial_\lambda H\ket{m}\bra{m} . \nonumber
\end{eqnarray}
Whereas this expression looks more similar to the full Kato AGP from Eq.~\eqref{eq:full_Kato_AGP} than its ground-state counterpart~\eqref{eq:GS_Kato_AGP}, it actually defines a ground-state AGP, as we now argue. 
First, notice that all diagonal elements of $A^\text{OH}_\text{GS}$ vanish as a result of the property $\tilde F(0)=0$; in particular, the Osborne-Hastings AGP respects the parallel transport gauge for the ground state: $(A_\text{GS}^\text{OH})^\text{gg}=0$.  
Next, consider the off-diagonal terms $n\neq m = 0$: because $E_n-E_0\geq\Delta$ is larger than the gap, $\tilde F_\text{OH}(E_n{-}E_0) {=} {-}(E_n{-}E_0)^{-1}$. As a result, we find that the Osborne-Hastings GS AGP agrees exactly with the GS Kato AGP in this sector: $(A_\text{GS}^\text{OH})^\text{eg}=(A_\text{GS}^\text{K})^\text{eg}$. 
From these two properties, it then follows that CD driving of the ground state using $A^\text{OH}_\text{GS}$ is equivalent to CD driving using $A^\text{K}_\text{GS}$. 

The difference between the Kato and Osborne-Hastings AGPs lies in the sector described by $A_\text{GS}^\text{ee}$. Osborne and Hastings' choice of the filter function is designed to guarantee a quasilocal (superpolynomial) spatial decay of the AGP $A^\text{OH}_\text{GS}$. This is essential for physical implementations of the operator. Intuitively, the filter cuts off energy scales smaller than the gap: notably, this eliminates the effect of the exponentially vanishing energy gaps between nearby eigenstates at high energy density, and produces a quasilocal AGP, formally proved using Lieb-Robinson bounds~\cite{osborne2007simulating,hastings2010locality}. In essence, $A^\text{OH}_\text{GS}$ makes use of the additional freedom in the excited-state sector to impose locality without changing the exact CD properties of the AGP w.r.t.~the ground state. 
By contrast, to the best of our knowledge there are no strict bounds that guarantee the (quasi-)locality of the Kato GS AGP $A_\text{GS}^\text{K}$ in general; although the ground state projector $\Pi_\text{GS}$ is a nonlocal operator in generic many-body systems, it is unclear under which conditions its commutator with its derivative~\eqref{eq:GS_HO_AGP} also lacks locality (e.g.,\ for the integrable transverse-field Ising chain, $A^\text{K}_\text{GS}$ is local when the system is gapped, cf.~Eq.~\eqref{eq:TFIC_AGP}, but this might be a consequence of the independent-two-level-system structure of this model).

Recall that the minimum norm condition for the ground-state AGP requires that $(A_\text{GS}^\text{K})^\text{ee}=0$ vanishes; 
at the same time, $(A_\text{GS}^\text{OH})^\text{ee}\neq 0$ clearly does not. 
It then follows that the Osborne-Hastings AGP trades norm for (quasi-)locality.

A natural question arises as to whether the transcritical CD driving defined in the main text with the help of the finite-speed parametrization, is feasible also with the Osborne-Hastings AGP. 
To answer it, consider the limit of vanishing gap $\Delta\to 0$, valid close to phase transition points, where the filter function becomes ineffective, and hence $\tilde F(\omega)=-\omega^{-1}$ for all $\omega$. In this limit, the OH gauge potential reduces to the \textit{full} Kato AGP,
\begin{equation}
    A^\text{OH}_\text{GS} \overset{\Delta\to 0}{\to} A^\text{K} ,
\end{equation}
rather than the GS AGP.
This follows directly from Eq.~\eqref{eq:full_Kato_AGP}, since $\tilde F_\text{OH}(\omega)\overset{\Delta\to 0}{\to} \tilde F(\omega)$.
It is then clear that close to phase boundaries, the Osborne-Hastings AGP suffers from the same problems as the full Kato AGP. In particular, its Hilbert-Schmidt norm density grows much larger compared to the ground-state Kato AGP $A^\text{K}_\text{GS}$, and eventually diverges. 

Nevertheless, since the OH GS-AGP agrees exactly with the Kato GS-AGP in its action on the ground state, the traversability properties of CD evolution with $A^\text{OH}_\text{GS}$ are the same as $A^\text{K}_\text{GS}$, provided the initial condition is the ground state in the bulk of the phase. The advantage of using $A^\text{OH}_\text{GS}$ is that it is provably quasilocal everywhere except at the phase transition point.

\end{appendix}

\else 
\fi 

\bibliography{src_latex/bibliography}
	
\end{document}